\documentclass[preprint]{JHEP3} 

\usepackage{epsfig}
\usepackage{amsmath}
\usepackage{amssymb,amsfonts}

\newcommand{\eqn}[1]{(\ref{#1})}

\newcommand{\eps}{\epsilon }

\newcommand{\Tr}{{\rm Tr}\ }
\newcommand{\Zint}{\mathbb{Z}}
\newcommand{\Real}{\mathbb{R}}
\def\Q{{\mathbb Q}}

\newcommand{\A}{{\cal A}}
\newcommand{\C}{{\cal C}}

\newcommand{\G}{{\cal G}}

\newcommand{\F}{{\cal F}}

\def\tr{{\rm tr}}

\def\a{\alpha}
\def\b{\beta}
\def\g{\gamma}
\def\l{\lambda}
\def\1d{{\dot 1}}
\def\2d{{\dot 2}}
\def\3d{{\dot 3}}
\def\ts{\textstyle}
\def\rt{{\rm t}}

\title{The Automorphic Membrane}

\author{Boris Pioline\\ LPTHE, Universit{\'e}s Paris 6 et 7 \\
Bo\^{\i}te 126, Tour 16, 1$^{\it er}$ {\'e}tage, 4 place Jussieu,\\ 
F-75252 Paris CEDEX 05, FRANCE\\
E-mail: \email{pioline@lpthe.jussieu.fr}}

\author{Andrew Waldron\\
Department of Mathematics,\\
One Shields Avenue,\\
University of California,\\
Davis, CA 95616, USA\\
Email: \email{wally@math.ucdavis.edu}}

\preprint{\hepth{0404018}v2\\
LPTHE/04-07\\
} 

\abstract{We present a 1-loop toroidal membrane winding
sum reproducing the conjectured $M$-theory, four-graviton,
eight derivative, $R^4$ amplitude. The $U$-duality and
toroidal membrane world-volume modular groups appear as
a Howe dual pair in a larger, exceptional, group. A detailed analysis
is carried out for $M$-theory compactified on a 3-torus, where
the  target-space $SL(3,\Zint)\times SL(2,\Zint)$ $U$-duality 
and $SL(3,\Zint)$ world-volume modular groups
are embedded in~$E_{6(6)}(\Zint)$. Unlike previous semi-classical
expansions, $U$-duality is built in manifestly and realized at the quantum
level thanks to Fourier invariance of cubic
characters. In addition to winding modes, a pair of new discrete, flux-like,
quantum numbers are necessary to ensure invariance
under the larger group. The action for these modes is of
Born-Infeld type, interpolating between standard Polyakov and
Nambu-Goto membrane actions. After integration over the membrane
moduli, we recover the known $R^4$ amplitude, including membrane
instantons. Divergences are disposed of by
trading the non-compact volume integration for a compact integral
over the two variables conjugate to the fluxes -- a constant term
computation in mathematical parlance.
As byproducts, we suggest that, in line with membrane/fivebrane duality,
the $E_6$ theta series also describes five-branes wrapped on $T^6$
in a manifestly U-duality invariant way. In addition we uncover a new action
of $E_6$ on ten dimensional pure spinors, which may
have implications for ten dimensional super Yang--Mills
theory. An extensive review of $SL(3)$~automorphic forms 
is included in an Appendix.}

\begin{document}

\def\Eis{\mbox{Eis}}

\def\wh{\widehat}
\def\wt{\widetilde}
\def\e{\varepsilon}
\def\bea{\begin{eqnarray}}
\def\eea{\end{eqnarray}}
\def\be{\begin{equation}}
\def\ee{\end{equation}}
\def\nn{\nonumber}
\def\ds{\partial\!\!\! /}
\def\dbs{\bar\partial\!\!\! /}
\def\uni{{\mathbb{1}}}
\def\A{{\cal A}}
\def\a{\alpha}
\def\b{\beta}
\def\D{\Delta}
\def\d{\delta}
\def\ve{\varepsilon}
\def\e{\epsilon}
\def\f{\varphi}
\def\F{\Phi}
\def\g{\gamma}
\def\G{\Gamma}\def\k{\kappa}
\def\l{\lambda}
\def\L{\Lambda}
\newcommand{\m}{\mu}
\newcommand{\n}{\nu}
\def\Om{\Omega}
\def\om{\omega}
\def\o{\omega}
\def\ph{\phi}
\def\r{\rho}
\def\s{\sigma}
\def\Si{\Sigma}
\def\t{\tau}
\def\th{\theta}
\def\Th{\Theta}
\def\w{\omega}
\def\x{\xi}
\def\z{\zeta}
\def\p{\partial}
\def\tS{\tilde S}
\def\Tr{\mbox{Tr}}
\newcommand{\ft}[2]{{\textstyle\frac{#1}{#2}}\,}
\newcommand{\XX}[2]{\{X^#1,X^#2\}}
\newcommand{\Wu}{\sqrt{\lambda^2+\tau_2\, \omega_2{}^2}}
\newcommand{\vt}{|\vec \tau|}
\newcommand{\vw}{|\vec w|}
\newcommand{\ra}{\rightarrow}
\def\C{{\cal C}}
\def\d{\partial}
\def\e{\varepsilon}
\def\nbar{\overline{\nu}}
\def\tbar{\overline{\tau}}
\def\Sl2Z{SL(2,{\mathbb Z})}
\newcommand{\comment}[1]{{\bf [#1]}}
\def\sss{\scriptscriptstyle}
\renewcommand{\ss}{\scriptstyle}

\tableofcontents

\section{Introduction}
Despite much evidence for its existence as a quantum theory,
a tractable microscopic definition of M-theory is still missing years
after the original conjecture~\cite{Townsend:1995kk,Witten:1995ex}. 
Various proposed definitions 
suffer either from a lack of computability
or ties  to specific backgrounds and energetics. Supermembranes 
remain one of the most promising candidates, because
they (i) imply the equations of motion of eleven-dimensional supergravity
as a consistency requirement 
(even at the classical level)~\cite{Bergshoeff:1987cm},  
(ii) reduce to the ordinary type II string by double 
reduction~\cite{Duff:1987bx}, and
(iii) are equivalent to a continuous version of M(atrix) theory in 
light-cone gauge~\cite{deWit:1988ig,deWit:1988ct,Banks:1996vh}.
However, the nonlinearities of the membrane world-volume theory and 
the lack of an obvious genus expansion
have so far stymied any attempt at direct quantization.

In a recent series of works, we have proposed to test the 
supermembrane $M$-theory hypothesis
in a simple setting avoiding the
usual quantization difficulties~\cite{Pioline:2001jn,Kazhdan:2001nx}. 
Our proposal is that
four graviton, eight derivative couplings in toroidally compactified
M-theory follow from a {\it one-loop BPS membrane amplitude}.
Here, ``one-loop'' refers to the membrane world-volume topology, 
namely a torus $T^3$, while ``BPS'' to the fact that only bosonic
and fermionic zero-modes of the embedding coordinate and the
world-volume metric contribute, fluctuations canceling by
virtue of supersymmetry.  

The basis for this  proposal can
be summarized as follows:
Supersymmetry requires that the $R^4$ amplitude receives 
corrections from BPS states only, 
as exemplified by the one-loop
computation of the $R^4$ amplitude in string theory, which indeed reduces 
to a sum of zero-mode  worldsheet instanton
configurations. In addition, the exact $R^4$ amplitude, 
determined on the basis of supersymmetry and 
$U$-duality~\cite{Green:1997tv,Green:1997di,
Kiritsis:1997em,Pioline:1997pu,Obers:1999um}, 
exhibits membrane instantons with only toroidal
topology, wrapped on $T^3$ subtorii of the target 
space $T^d$ \footnote{Membrane instanton corrections in less
supersymmetric settings have been discussed in \cite{Harvey:1999as}.}.
Therefore, we expect that only toroidal membrane topologies $T^3$
contribute to this amplitude. Since $U$-duality relates membrane
instantons to perturbative contributions, a treatment of 
membrane instantons that maintains 
{\it manifest $U$-duality} symmetry, will necessarily
reproduce the complete $R^4$ amplitude, 
including its compactification-independent part.

In this Article, we demonstrate that our proposal~\cite{Pioline:2001jn}, 
does yield the correct $R^4$ amplitude, in the simplest non-trivial
case of M-theory compactified on a 3-torus. Instead of quantizing 
the classical membrane action,, we assume that the
result of this quantization will reduce to a sum over discrete
zero-mode configurations, invariant under all quantum symmetries. 
Invariance under the U-duality group $SL(3,\Zint)\times SL(2,\Zint)$
and the world-volume modular group $SL(3,\Zint)$ is then shown
to require a larger symmetry group, $E_{6(6)}(\Zint)$, under which
the partition function is automorphic.
In the process, we discover new discrete, flux-like, degrees 
of freedom on the membrane worldvolume, necessary to ensure
U-duality invariance. We are also lead to a new prescription for
dealing with the divergence associated to the membrane volume,
which makes a crucial use of these new degrees of freedom. 
Our calculations are mathematically rigorous only in the degenerate
sectors, corresponding to Kaluza-Klein contributions, the proof
in the non-degenerate, membrane instanton sector falls short
at a technical difficulty, for which we can only suggest a resolution.

This Article is organized as follows: In Section~\ref{R4review}, we outline
the general strategy proposed in~\cite{Pioline:2001jn}, and review the exact
$R^4$ amplitude for M-theory compactified on a 3-torus,  
obtained in~\cite{Kiritsis:1997em}
on the basis of supersymmetry and $U$-duality. In Section~\ref{E6theta}, 
we recall the construction of the $E_6$ theta series from
our earlier work~\cite{Kazhdan:2001nx} and identify
physical parameters $(\gamma_{AB}; g_{MN}, C_{MNP})$ with 
those entering the exceptional theta series. 
In Section~\ref{wvintegrals}, we  
carry out the integration over the shape moduli of the world-volume $T^3$
and show how to produce the $U$-duality $R^4$ amplitude from a constant 
term integration which regulates the membrane world-volume integration.
In Section~\ref{highertorii}, 
we indicate how these results generalize to higher
torii. We also point out the possible relevance of the $E_6$
theta series for five-branes compactified on $T^6$, which indicates
that membrane/five-brane duality may be contained in our framework.
Not unrelatedly, we also uncover a new action of $E_6$ on 
pure spinors in ten dimensions (first announced 
in~\cite{leshouches}). It would be interesting to understand the 
implications for ten dimensional super Yang--Mills theory. 
Our conclusions, including various caveats and speculations are
presented in Section~\label{discussion}. Finally, Appendix A gives
more details on the $E-6$ theta series, Appendix B describes
relevant aspects of $SL(3)$ automorphic forms which may usefully
supplement our recent review~\cite{leshouches}, and Appendix
C is a sample membrane volume integral computation, illustrating
the typical divergence of such integrals.

\section{$R^4$ couplings, membranes and theta series: a review}
\label{R4review}

In this section, we outline the general approach, first proposed
in~\cite{Pioline:2001jn}. We then review the existing knowledge
about $R^4$ couplings in 8 dimensions, which we shall attempt
to derive from the membrane prospective.

\subsection{$R^4$ couplings and theta correspondences}

To summarize the salient steps of our construction, recall first that
the analogous one-loop computation in type II string theory
compactified on $T^d$
yields a partition function for  string winding modes
invariant under the world-volume modular group $SL(2,\Zint)$ times
the $T$-duality group $SO(d,d,\Zint)$. The $R^4$ amplitude is then 
obtained by  
integrating over the fundamental domain of the toroidal lattice moduli space 
$SL(2,\Real)/SO(2)$
\be
\label{zstr}
f_{R^4}^{\rm 1-loop}=
\int_{\!\!\!\!\!\!\!_{{}_{SO(2)}\backslash^{SL(2,\Real)}/_{SL(2,\Zint)}}} 
\!\!\!\!\!d\gamma\ \; Z_{\rm str}(\gamma_{\a\b}; g_{\mu\nu},B_{\mu\nu})\, .
\ee
Here $\gamma_{\a\b}$ is the unit volume 
constant metric on the world-volume 2-torus while
$g_{\mu\nu}$ and $B_{\mu\nu}$ are the constant target space $d$-torus 
metric and 2-form gauge
field.
Correspondingly, the one-loop BPS membrane amplitude  
for M-theory compactified on $T^d$
is a partition function for membrane winding modes, invariant 
under the modular group of the 3-torus $SL(3,\Zint)$ times
the $U$-duality group $E_d(\Zint)$. The $M$-theory 
four graviton, eight-derivative amplitude ought emerge after 
integrating over the fundamental
domain of the moduli space
of constant metrics on the toroidal world-volume:
\be
\label{zmem}
f_{R^4}^{\rm exact}
\stackrel{?}{=}
\int_{\!\!\!_{{}_{SO(3)}\backslash^{Gl(3,\Real)}/_{SL(3,\Zint)}}} 
\!\!d\gamma\; Z_{\rm mem}(\gamma_{AB}; g_{MN},C_{MNP})\, .
\ee
Here $\gamma_{AB}$ is the constant metric on the world-volume 3-torus
and $g_{MN}$ and $C_{MNR}$ the constant toroidal target-space
metric and 3-form gauge field.
In contrast to string amplitude~\eqn{zstr}, we expect to integrate over
volume of the metric $\gamma_{AB}$  
because the membrane world-volume theory is not conformally invariant.
We return to this important point later.

As a preliminary study, a semi-classical Polyakov quantization of
the BPS membrane, based on a constant
instanton summation measure, was performed in~\cite{Pioline:2001jn}.
We found the correct instantonic saddle points: Euclidean
membranes wrapping all subtorii $T^3$ of the target space $T^d$, but
with an incorrect summation measure, incompatible with
U-duality. Similarly, the Hamiltonian interpretation of the same result gave 
a consistent spectrum of BPS states running inside the loop, 
but with the wrong multiplicities. This result is hardly surprising,
since the membrane world-volume theory exhibits $U$-duality 
symmetry at the classical level~\cite{Duff:1990hn,Lukas:1997jk},
but quantum membrane self-interactions are expected to depend non-trivially
on the instanton number. This was confirmed in a  $U(N)$ matrix model
computation in \cite{Sugino:2001iq}, who found the correct
instanton summation measure, in agreement with expectations from U-duality. 

This
 failure of semi-classical quantization suggests a  different
approach: $U$-duality invariance should be manifestly built in  
from the beginning. To that end,
observe that the string theory partition function is 
simply a standard Gaussian
theta series, invariant under the symplectic group $Sp(2d,\Zint)$,
which contains $SL(2,\Zint)\times SO(d,d,\Zint)$ as commuting
subgroups. The string theory moduli $g_{\mu\nu}, B_{\mu\nu}$ live
on a slice of the symplectic 
period matrix moduli space according to the decomposition
\be 
\Big(SL(2,\Real)/SO(2)\Big)\times 
\Big(SO(d,d,\Real)/[SO(d)\times SO(d)]\Big)
\subset
Sp(2d)/U(d)\, , 
\ee
which is clearly 
preserved under the  discrete $SL(2,\Zint)\times SO(d,d,\Zint)$ 
$T$-duality subgroup\footnote{For $d=2$, the group $Sp(4)$ contain
extra discrete generators which preserve this slice, which in 
particular can exchange the world-sheet and target space complex 
structures~\cite{Verlinde}.}.
More generally, the string winding mode partition function
gives a correspondence between 
automorphic forms of the modular group $SL(2,\Zint)$ and the
$T$-duality group $SO(d,d,\Zint)$, by integrating 
with respect to the fundamental domain of either of the two factors.
By analogy, we propose that the partition function of the
membrane winding modes should be a theta series for a larger group
including $SL(3,\Zint)\times E_d(\Zint)$ as commuting subgroups.
Specifically, we consider 
the following candidates
\be
\label{embeddings}
\begin{array}{cccc}
d=3& \quad Gl(3)\times [ SL(2)\times SL(3) ] &\subset& E_6\\
d=4& Gl(3)\times [ SL(5) ]&\subset& E_7\\
d=5& Gl(3)\times [ SO(5,5)] &\subset& E_8\, .
\end{array}
\ee
Here the $Gl(3)$ factor is to be interpreted as the world-volume
modular group of a toroidal
membrane (including the $\Real^+$ volume factor) while the bracketed
factor is the $U$-duality group $E_d$ of $d$-toroidally compactified
$M$-theory. The choice of the overarching group $E_{6,7,8}$ is the
most economic guess, and will be justified {\it a posteriori} below. 
A prime reason to consider these groups is the fact that their minimal
representation naturally involves cubic characters, which are expected
when dealing with membrane Chern--Simons couplings.

According to our proposal,
BPS supermembrane quantization therefore 
amounts to the construction of theta
series for the exceptional groups $E_{6,7,8}$. The existence of
exceptional theta series is
known to mathematicians, but explicit expressions were unavailable
until recently. 
In an earlier Article we have explicitly
computed the generic summand in the theta series for any simply laced
group $G$ using techniques of representation theory~\cite{Kazhdan:2001nx}.
The explicit summation measure and non-generic 
degenerate contributions have  in turn
been computed in~\cite{KazPol}. An important feature of exceptional
theta series is that invariance under the arithmetic subgroup $G(\Zint)$
does not reduce to the usual Poisson resummation of Gaussian sums, but
instead involves Fourier invariant cubic characters. This meshes nicely
with the inherently cubic membrane couplings
to the background 3-form $C_{MNR}$. Another important result 
from~\cite{Kazhdan:2001nx}, 
is the requirement of extra flux-like quantum numbers
over and above membrane winding numbers 
to realize the larger exceptional symmetries.

In this work, we concentrate on the $d=3$ case, which has all the features
of interest including membrane instantons, and carry out the integration
in~\eqref{zmem} using for $Z_{\rm mem}$ the theta series of $E_6$.

\subsection{Exact $R^4$ amplitude in 8 dimensions}

The four-graviton, eight-derivative amplitude in M-theory has received
much attention in recent years, since it is determined completely
non-perturbatively on the basis of supersymmetry and $U$-duality. 
The original conjecture by Green and Gutperle for type~IIB 
string theory in ten dimensions~\cite{Green:1997tv}
was later generalized to compactification 
on higher-dimensional tori $T^d$ in~\cite{Kiritsis:1997em,Green:1997di,
Pioline:1997pu,Obers:1999um}. 
For the case $d=3$ of
main interest in this Article, the result proposed 
in~\cite{Kiritsis:1997em} reads
\be
f_{R^4} = {\widehat E}^{SL(3)}_{{\bf 3},3/2}(G) 
+ {\widehat E}^{SL(2)}_{{\bf 2},1}(\tau)
\label{fr4}
\ee
where the two terms are Eisenstein series for the two factors of the
$U$-duality group $SL(3,\Zint) \times SL(2,\Zint)$,
namely
\be
\label{eis32}
E^{SL(3)}_{{\bf 3},s} = \sum_{m^M\in \Zint^3 \backslash \{ 0 \} }
\left[ \frac{V^{2/3}}{m^M G_{MN} m^N} \right]^s \ ,\quad
E^{SL(2)}_{{\bf 2},s} = \sum_{(m,n)\in \Zint^2 \backslash \{ 0 \} }
\left[ \frac{\tau_2}{|m+n\tau|^2} \right]^s  \ ,\quad
\ee
where $G_{MN}$ is the constant  metric on the target
space torus $T^3$ with volume $V=\det^{1/2} G_{M N}$, 
and $\tau$ is the volume modulus $\tau=C+i \frac{V}{l_M^3}$
complexified by the constant 3-form gauge field
$C=C_{123}$ on $T^3$. The moduli take values in the 
product of symmetric spaces
\be
(G_{MN},\tau) \in  \left[ SO(3)  \backslash SL(3,\Real) \slash SL(3,\Zint)
\right] \times   \left[ SO(2) \backslash SL(2,\Real) \slash
 SL(2,\Zint)  \right]
\, ,
\ee
and transform under the left action of the $U$-duality group. 
The hat appearing on $E$ denotes the fact
that the automorphic forms appearing in~\eqref{fr4} are really the
finite terms following the pole at $s=3/2$ (respectively $s=1$),
\be
\label{spole}
E^{SL(3)}_{{\bf 3},s} = \frac{2\pi}{s-3/2}\ +\  
{\widehat E}^{SL(3)}_{{\bf 3},3/2}\ +\ {\cal O}(s-3/2)\ ,\quad
E^{SL(2)}_{{\bf 2},s} = \frac{\pi}{s-1}\ +\  
{\widehat E}^{SL(2)}_{{\bf 2},1}\ +\ {\cal O}(s-1)\, .
\ee
The conjecture~\eqref{fr4} has been checked in many different ways:
\begin{enumerate}
\item The original motivation came from a perturbative analysis 
in type IIA string theory compactified on $T^2$. {}From this viewpoint,
the $3\times 3$ matrix $G_{MN}$ encodes the string coupling constant
$g_s$ and the complex structure $U$ of the 2-torus, whereas the
complex modulus $\tau$ corresponds to the complexified
K\"ahler class $T$ of the 2-torus. Upon expansion around the cusp at $g_s=0$, 
the result~\eqref{fr4} reproduces the tree-level and one-loop contributions
in type IIA compactified on $T^2$, together with an infinite set of
non-perturbative effects attributed to Euclidean D0-branes winding around
the torus $T^2$:
\be
f_{R^4} = \left[ \frac{2\zeta(3) V}{g_s^2} 
-2\pi \log U_2 ^4|\eta(U)|^2 + f_{R^4}^{D0} \right]
+ \left[ -2\pi \log T_2 ^4|\eta(T)|^2 \right]
\ee
Here the two terms in brackets correspond to the contribution of the
first and second term in~\eqref{fr4}, respectively. The summation measure
for D0-instantons can be easily extracted from this result,
and was successfully rederived from a matrix model 
computation in~\cite{matrixD0comp}.
The second term arises
purely at one-loop, and includes the effect of perturbative worldsheet
instantons. Actually, the two terms above contribute to different
kinematical structures $(t_8 t_8$~$\pm$~$\epsilon_8\epsilon_8) R^4$, which
we will not distinguish (they are identical in dimension $D$~$\leq$~$7$).
The vanishing of the $R^4$ amplitude at two loops implied by~\eqref{fr4}
has been recently confirmed~\cite{Iengo:2002pr} (see 
also~\cite{D'Hoker:2002gw} for a review of 
recent advances on two-loop technology).

\item The first term in~\eqref{fr4} was derived up to an infinite additive 
constant from a one-loop computation
in eleven-dimensional supergravity compactified on a two torus for which
the integer summation variables $m^M$ are Kaluza-Klein 
momenta of the graviton, or rather their canonical 
conjugates~\cite{Green:1997as}. 
(This computation missed the second term, which can
be attributed to membrane instantons wrapping $T^3$.) Indeed,
expanding~\eqn{fr4}  
around the cusp at $V \to \infty$, one obtains
\bea
f_{R^4}&=&\frac{2\pi^2 V}{3 l_M^3} 
         +\sum_{m^M\!\in {\mathbb Z}^3\setminus\{0\}}
          \frac{V}{\Big[m^{[1]} m_{[1]} \Big]^{3/2}}
-2\pi \log(V/l_M^3) \nn\\
       &+& \pi V \!\!\! \sum_{m^{[3]} \in \Zint\setminus\{0\}}
           \frac{\mu(m^{[3]})}{\sqrt{\left|m^{[3]}m_{[3]}\right|}} \
           \exp\Big(-\frac{2\pi}{l_M^3} 
                     \sqrt{\left|m^{[3]}m_{[3]}\right| }\ +\ 
           2\pi i m^{[3]} C_{[3]}\Big)\, .\nn\\
\label{instantons}
\eea
The single summation integer $m^{[3]}$ can be viewed as a target space
three-form\footnote{\label{soul}So $m^{[3]}=m^{123}$ and $m^{[3]}m_{[3]}=V^2(m^{123})^2$
where the volume factors come from three contractions with the target
space metric $G^{MN}$. } $m^{MNP}$ which
counts  target 3-torus wrappings of an M2-brane
with tension~$1/l_M^3$.
This is the main reason to expect 
that a one-loop supermembrane computation should
reproduce membrane instantons and hence the full $R^4$ couplings
if $U$-duality can be maintained.
In this Article, we search for a derivation of this result in
terms of fundamental super membrane excitations so that wrappings
are expressed in terms of winding numbers
\be
m^{MNP}=\frac{1}{3!}\
\epsilon^{ABC}\ Z_A^MZ_B^NZ_C^P\, .
\label{winding_numbers}
\ee
The integer-valued matrix $Z_A^M$
counts windings of the $A$th world-volume cycle
about the $M$th target space one. 
The summation measure $\mu(m^{[3]})$ for membrane instantons on $T^d$ 
is easily extracted from the $SL(2)$ weight 1 Eisenstein series,
and reads
\be
\label{measure}
\mu(m^{[3]})=\sum_{n|m^{[3]}}n\, .
\ee
A semi-classical supermembrane computation~\cite{Pioline:2001jn} already yields
the correct $U$-duality invariant exponent in~\eqn{instantons}
but does not correctly predict the counting of states given by $\mu(m^{[3]})$.
It is an important challenge to rederive 
the result~\eqn{instantons} including the correct measure factor 
from the membrane 
theory.

\item It was shown at the linearized level for
M-theory on $d=3$~\cite{linearMSUSY} or at the non-linear level for type 
IIB~\cite{nonlinIIBSUSY} that supersymmetry requires the $R^4$ amplitude to be
an eigenmode of the Laplacian on the moduli space with a specific eigenvalue:
\be
\label{dso}
\Delta^{(2)}_{SO(3)\backslash SL(3)} f_{R^4} =
\Delta^{(2)}_{SO(2) \backslash SL(2)} f_{R^4} =
 0
\ee
in the $d=3$ case. It is also possible to check that the $SL(3)$ part
is annihilated by the cubic Casimir $\Delta^{(3)}_{SO(3)\backslash SL(3)}$,
defined below in~\eqref{casimirs}. In fact, the analysis 
in~\cite{linearMSUSY,nonlinIIBSUSY} was not sensitive to possible 
holomorphic anomalies, and indeed, due to the subtraction of the
pole in~\eqref{spole}, the right-hand side of~\eqref{dso} picks up
a non-vanishing constant value, $2\pi$. This can be viewed as the
result from logarithmic infrared divergences, as expected for $R^4$ couplings
in 8 dimensions, in close analogy with gauge couplings in 4 
dimensions~\cite{Dixon:1990pc}\footnote{Recall that in the terminology 
of~\cite{Dixon:1990pc}, the holomorphic anomaly originates from the 
degenerate orbit, {\it i.e.}, the contribution of Kaluza-Klein gravitons.}.
Under mild assumptions on the
behavior at small coupling, 
equation~\eqref{dso}together with U-duality invariance in fact imply
the correctness of~\eqref{fr4}.
\end{enumerate}
The result~\eqref{fr4} is therefore on a very firm footing,
and is one of the few exact non-perturbative M-theoretic results available. 
Its derivation from first principles is therefore a central problem of 
this field, to which we now turn.

\section{$E_6$ minimal representation and theta series}
\label{E6theta}

As discussed in the Introduction, our contention is that the
exact $R^4$ amplitude~\eqref{fr4} follows from a one-loop computation 
in the BPS membrane theory, whose partition function
is provided by the $E_6$ theta series. In this Section we 
recall the motivation for this claim, review the construction of the
$E_6$ theta series, and identify the physical parameters inside
the moduli space of the $E_6$ theta series.

\subsection{{}From modular and $U$-duality invariance to $E_6$}

We posit that the correlator of four graviton vertex operators
on a supermembrane of topology $T^3$, at leading order in momenta,
reduces to the partition function for the membrane winding modes, 
except for a summation measure which incorporates the effects
of non-Abelian interactions for multiply wrapped membranes. The
partition function of these winding modes should furthermore be
invariant under $SL(3,\Zint) \times SL(2,\Zint)$ $U$-duality, together
with $SL(3,\Zint)$ modular transformations. Invariance
under the two $SL(3,\Zint)$ factors is automatic as soon as the membrane
theory is invariant under world-volume and target space diffeomorphisms.
Invariance under the $SL(2,\Zint)$ factor is 
a highly
non-trivial statement from the point of view of the membrane, although under
double reduction it amounts to the $T$-duality symmetry of the type II string.
A natural way to realize this symmetry is to require
invariance under the larger group $E_6$, through the two-stage
decomposition
\be
SL(3)\times \left( SL(2)\times \Real^+ \right) 
\times SL(3) \subset SL(3)^3 \subset E_6
\ee
Other possibilities exist, but only this one seems to lead to a natural
membrane interpretation. One should therefore construct an automorphic
theta series for $E_6$. This has been undertaken
in~\cite{Kazhdan:2001nx,KazPol}, which we  
review here for completeness: 

\subsection{Theta series, representations and spherical vectors}

Given a group $G$, a theta series is constructed as follows
(see {\it e.g.}~\cite{leshouches} for a more complete discussion): Let $\rho$
be a representation of $G$ acting on functions of some space $X$. Let
$f$ be a function invariant under the action of the maximal compact
subgroup $K\subset G$,
\be
\rho(k)f=f\, ,\qquad\forall k\in K\, .
\ee
For so called spherical representations, this function is unique and
known as the spherical vector.  Let  $\delta$ be  a distribution 
in the dual space of  $X$, invariant 
under an arithmetic subgroup $G(\Zint)\subset G$. Then the  series
\be
\theta(g)=\langle\delta,\rho(g)f\rangle\, ,\label{formula}
\ee
is a function on $G/K$, invariant under right action 
by $G(\Zint)$, hence an automorphic form\footnote{To be precise, automorphic
forms are usually required to be eigenfunctions of the Laplacian and
higher order invariant differential operators associated with
the Casimirs of $G$. This may be achieved by requiring irreducibility
of the representation $\rho$.} for $G(\Zint)$. 
The term ``theta series'' is usually reserved for the case
when $\rho$ is the minimal representation, {\it i.e.} the one of smallest
functional dimension. In particular, 
the minimal representation carries no free parameter (excepting 
$A_n$, which admits a one-parameter family of minimal
representations, associated to the homogeneity degree $s$ 
in Eisenstein series such as~\eqref{eis32}). 
This is a desirable feature for applications to 
supermembrane physics.
The minimal representation was constructed explicitly for
all simply laced groups in~\cite{KazSav}, based on the quantization of
maximally nilpotent coadjoint orbits. The spherical vector was
computed in our earlier work~\cite{Kazhdan:2001nx}, and the 
invariant distribution
$\delta$ was obtained recently by adelic methods~\cite{KazPol}.
For example, when $G=Sp(2,\Zint)$, 
one recovers the standard Gaussian theta series, 
as follows: The minimal representation is the metaplectic 
representation acting 
on $X=L_2(\Real)$ with generators $\p_x^2, x^2$ and $
x\p_x +\p_x x$; the (quasi)spherical vector $f$ is the Gaussian
$f=e^{- x^2}$, and $\delta$ is the comb distribution
$\delta_{\Zint}$, invariant under integer shifts and Fourier transform.
The latter can be viewed adelically as the product over all primes $p$ of
the spherical vector $f_p$ over the $p$-adic field $\mathbb{Q}_p$,
which is simply the unit function on the $p$-adic integers, invariant
under integer shifts and Fourier transform.

\subsection{Minimal representation of $E_6$}
\label{minrep}

We now turn to details of the $E_6$
theta series. Representations of non-compact groups are generally
obtained by quantizing the coadjoint orbit of an element $e$ in $G^*$. 
The representation of smallest dimension arises upon choosing
a non-diagonalizable element $e$ of maximal nilpotency, which
can always be conjugated into the lowest root $E_{-\omega}$.
The generators $E_{-\omega}, E_{\omega}$ and $H_\om = [E_\om,E_{-\om}]$
form an $SL(2)$ subalgebra, with maximal commuting algebra $SL(6)$
in $E_6$, 
\bea
E_6&\supset& SL(2)\times SL(6) \nn\\
78&=&(3,1)\oplus(2,20)\oplus(1,35)\, .
\eea
The Cartan generator  $H_{\omega}$ of this $SL(2)$ subalgebra
grades $G$ into~5~subspaces which form representations
of the commutant $SL(6)$:
\bea
\begin{array}{|c|ccccc|}
\hline
\mbox{$\ \ H_\omega\ $ charge \ \ }&2&1&0&-1&-2\\
\hline 
\mbox{$SL(6)$ irreps}&1&20&1+35&20&1\\
\hline 
\mbox{generators}&\  E_\omega\  &\  \{E_\beta,E_\gamma\}\  &
\  \{H_\omega,H_\beta,H_\gamma,H_\alpha,E_{\pm \alpha}\}\  &
\  \{E_{-\beta},E_{-\gamma}\}\  &\  E_{-\omega}\  \\
\hline
\end{array} \nn\\ \nn\\ 
\eea
The stabilizer $S$ of $E_{-\omega}$ is given by the grade~-2 and grade~-1
subspace, together with the non-singlet part of the grade~0~subspace. 
Therefore, the
coadjoint orbit ${\cal O}=S\backslash G$ of $E_{-\omega}$ can be
parameterized by the 
orthogonal complement to the stabilizer $S$, 
namely the grade~1 and grade~2 spaces,
together with the singlet in the grade~0 space. It carries the canonical
Kirillov-Kostant symplectic structure and admits a
right action of the group~$G$. The action of~$G$ can thus
be represented  in terms of canonical generators acting by Poisson brackets.
For~$E_6$, we then have a representation on functions of a 22~dimensional,
classical, phase space~${\cal O}$. 

The minimal representation of $E_6$ acts on functions of half as many variables
and is obtained by quantization of the classical system. The first step is
to choose a polarization or Lagrangian subspace of ${\cal O}$, {\it i.e.}: 
split the 22~coordinates
into~11 positions and momenta. This is easily done by noting that
the grade 1 subspace commutes to $E_{\om}$ and hence forms a Heisenberg 
subalgebra.
A particular choice of polarization is to break the symmetry $SL(6)$
to a subgroup $H_0=SL(3)\times SL(3)$ realized linearly on positions, 
which decomposes the grade 1 subspace as
$20=1+(3,3)+(3,3)+1$. We choose as positions one of the two copies
of $1+(3,3)$ and call $(E_{\gamma_0},E_{\gamma_A^M})$ 
position operators with  $(E_{\beta_0},E_{\beta_M^A})$ being
their conjugate momenta. An extra position variable $y$
corresponding to $E_{\om}$ plays the r\^ole of $\hbar$.
These generators are represented in the usual way
\be
\begin{array}{lcl}
&E_\omega=iy\, ,&\\
 \\
E_{\beta_0}=y\ \d_{x_0}\, ,&& E_{\gamma_0}=ix_0\, ,\\
E_{\beta_M^A}=y\ \d_{Z_A^M}\, ,&& E_{\gamma_A^M}=iZ_A^M\, .\\
\end{array}
\ee
The remaining generators may be obtained using Weyl reflections
as explained in~\cite{KazSav,Kazhdan:2001nx}, and are 
displayed in Figure~\ref{E6_rep}. Pertinently, the generator
for the negative root $-\beta_0$ involves the cubic $H_0$ invariant
$I_3\equiv \det Z$,
\be
E_{-\beta_0}=-x_0\d_y-\frac{\textstyle i\det Z}{y^2}\, .
\ee
Functions invariant under the maximal compact subgroup $Usp(4)$ of
$E_6$ must be annihilated by the compact generator
\be
K_{\beta_0}=E_{\beta_0}+E_{-\beta_0}=
y\ \d_{x_0}-x_0\d_y-\frac{\textstyle i\det Z}{y^2}\, .
\ee
Recognizing the first two terms as an $SO(2)$ rotation generator,
this restricts the $Usp(4)$ invariant functional space to
\be
f(y,x_0,Z_A^M)=\exp\Big(i\frac{x_0\det Z}{y(y^2+x_0^2)}\Big)\, 
\hat f(y^2+x_0^2,Z_A^M)
\, .
\ee
Automorphic $E_6$ invariance in this polarization of the minimal
representation relies on (Fourier) invariance of the cubic character
$\exp(i \det(Z)/x_0)$ (these are discussed
further in~\cite{Pioline:2003uk}). 
Furthermore, in this scheme, the shifts of the target space 
three-form $C$ can be identified with the action of 
the generator $E_{-\beta_0}=y\p_{x_0}$, whose effect is to 
shift  $x_0\rightarrow x_0+Cy$. The spherical functions thus
couple to the $C$ field by a {\it cubic} phase
\be
\exp\Big( i\ \frac{C \det Z}{y^2+[x_0+Cy]^2}\Big)\, .
\ee
reminiscent of the Chern--Simons coupling $\exp( i C \det Z)$
expected for a membrane. This cubic coupling is one of the main 
reasons to believe that $E_6$ can appear as an overarching group
for the membrane on $T^3$. The appearance of additional variables
$(y,x_0)$ in the denominator is a new feature predicted by this
$E_6$ symmetry.

To summarize, $E_6$ is
represented on functions of 11 variables $(y,x_0,Z_A^M)$
with $a=1,2,3$ and $A={\1d,\2d,\3d}$. The generators of the 
$SL(3)\times SL(3)$ 
subgroup act linearly  by matrix multiplication
on $Z=(Z_A^M)$ from the left and right leaving $y$ and $x_0$
invariant, while all remaining generators
act non-linearly.
The physical interpretation of these 11 variables will be discussed below.

\begin{figure}
{\small
\be
\begin{array}{lcl}
E_\omega&=&iy\\
 \\ \\
E_{\beta_0}&=&y\ \d_{x_0}\\
E_{\beta^A_M}&=&y\ \d_{Z_A^M}\\
E_{\gamma_0}&=&ix_0\\
E_{\gamma_A^M}&=&iZ_A^M\\
 \\ \\
L_A^B&=&-Z_A^M\d_{Z_B^M}\;=\;-E_{-\a_B^A}\qquad  SL(3)_L\\
R^M_N&=&-Z_A^M\d_{Z_A^N}\;=\;-E_{-\a^N_M}\qquad SL(3)_R\\
 \\
E_{\a^A_M}&=&-x_0\d_{Z_A^M}+\frac{i}{2y}\ \e^{ABC}\e_{MNR}Z_B^NZ_C^R\\
E_{-\a_A^M}&=&Z_A^M\d_{x_0}-\frac{iy}{2}\e_{ABC}\e^{MNR}\d_{Z_B^N}\d_{Z_C^R}
\\  \\
H_{\a_A^M}&=&-x_0\d_{x_0}+Z_A^M\d_{Z_A^M}+
(1-\delta_{AB})(1-\delta_{MN})Z_B^N\d_{Z_B^N}+2\quad\mbox{(no sum on $A$, $M$)}\\
H_{\beta_0}&=&-y\d_y+x_0\d_{x_0}\\
H_{\omega}&=&-2y\d_y-x_0\d_{x_0}-Z\cdot \d_Z-6\\
 \\
E_{-\beta_0}&=&-x_0\d_y-\frac{i\det Z}{y^2}\\
E_{-\beta^A_M}&=&Z_A^M\d_y+\frac i2\ x_0\
\e_{ABC}\e^{MNR}\ \d_{Z_B^N}\d_{Z_C^R}
+\; \frac{1}{y}\ (Z_A^M\ [Z\cdot \d_Z+2]-Z_A^NZ_B^M\d_{Z_B^N}) 
\\
E_{-\gamma_0}&=&-y\det[\d_Z]-i(y\d_y+x_0\d_{x_0}+Z\cdot \d_Z+6)\d_{x_0}\\
E_{-\gamma_A^M}&=&i(y\d_y+x_0\d_{x_0}+4)\d_{Z_A^M}
+iZ_A^N\d_{Z_A^N}\d_{Z_B^M}
+\frac{1}{2y}\ \e^{ABC}\e_{MNR}Z_B^NZ_C^R
\\
\\
E_{-\omega}&=&-i(y\d_y+x_0\d_{x_0}+Z\cdot \d_Z+6)\d_y+
x_0\det[\d_Z]\\&&
-\frac{i}{y}\ \Big(2Z\cdot \d_Z+\frac12(Z\cdot \d_Z)^2
-\frac12Z_A^MZ_B^N\d_{Z_B^N}\d_{Z_A^N}+6\Big)
+\frac{\det(Z)}{y^2}\ \d_{x_0}
\\
\end{array}
\ee}
\caption{Infinitesimal generators for the $E_6$ minimal representation.}
\label{E6_rep}
\end{figure}

\subsection{$E_6$ spherical vector}

The spherical vector $f_{E_6}$ is invariant under the maximal
compact subgroup $Usp(4)$ 
of $E_6$ generated 
by $K_\delta=E_\delta+E_{-\delta}$ for any
positive root $\delta$. So determining the spherical vector
amounts to solving a complicated system partial differential equations 
$K_\delta f_{E_6}=0$. Invariance under the maximal compact subgroup 
of the linearly acting $SL(3)\times SL(3)$ implies that $f_{E_6}$
is a function of the quadratic, cubic and quartic invariants
$\tr Z^\rt  Z,\, \det(Z),$ and  $\tr (Z^\rt  Z)^2$. Invariance under
the remaining compact generators fixes this function to 
be~\cite{Kazhdan:2001nx}
\be
f_{E_6}=\frac{\exp(-S_1-iS_2)}{(y^2+x_0^2)\ S_1}\, ,
\label{spherical_E6}
\ee
where
\be
S_1=\frac{\sqrt{\det[ Z Z^\rt +(y^2+x_0^2)\mathbb{I}\ ]}}{y^2+x_0^2} \ ,\qquad
S_2=-\frac{x_0\det(Z)}{y(y^2+x_0^2)}\, .
\label{action}
\ee
The exponential weight $S_1+iS_2$ should be thought of as the classical action
of a membrane with quantum numbers $y,x_0,Z^M_A$, at the origin of the moduli
space $E_6/Usp(4)\supset [SL(3,\Real)/SO(3)]^3$. In the next Section
we explain how to couple the theory to world-volume and target-space moduli.
At this point, we note that the integers $Z^M_A$ are naturally interpreted
as windings arising from the zero-modes of transverse membrane coordinates:
\be
X^M(\sigma^A)=Z^M_A\sigma^A+\cdots\, , \qquad Z^M_A=\d_A X^M\, .
\ee
The action $S_1$ is a Born-Infeld
type-generalization of the Polyakov membrane action.

The additional quantum numbers $y,x_0$ are necessary for manifest
$U$-duality but cannot correspond to 
propagating membrane world-volume degrees of freedom
which have already been accounted for by the windings $Z_A^M$.
We propose, therefore,
that they instead 
correspond to field strengths of a pair of two-form gauge fields 
$B_{AB}$ and $B_{AB,MNR}$ (transforming as  target space scalar and
three-form densities), which have no
propagating degrees of freedom in 3 dimensions, but whose
field strengths take only quantized values.

Clearly if correct, our proposal, based on maintaining $U$-duality, 
provides a detailed microscope for examining fundamental membrane
$M$-theory excitations.
In particular, it would be interesting to extend $U$-duality invariance
of the classical zero-mode action to quantum fluctuations.
A remarkable feature of the $E_6$ spherical vector~\eqref{spherical_E6} 
is the simple exponential
(corresponding to the Bessel function of index~$1/2$) which
receives no quantum corrections about the classical action. 
In contrast, higher $E_n$ cases  involve genuine, ``quantum corrected'',
Bessel functions of~$S_1$.

\subsection{Identification of physical parameters}
\label{paramsect}

We now present the decomposition of the 
extended duality group $E_6$ into world-volume modular and
target space $U$-duality groups.
Examining the $E_6$ Dynkin diagram in Figure~\ref{E6},
we identify three commuting $SL(3)$ subgroups with positive roots
\be
\label{sl3s}
\Delta^+_{SL(3)_{L}}=\{\a_{12},\a_{23},\a_{13}\}\, ,\quad\!\! 
\Delta^+_{SL(3)_{R}}=\{\a_{\1d\2d},\a_{{\2d\3d}},\a_{\1d\3d}\}\, ,\quad\!\!
\Delta^+_{{SL(3)}_{NL}}=\{\beta_0,-\omega,-\gamma_0\}\, .
\ee 
The first two factors $SL(3)_{L}$ and $SL(3)_R$ act linearly on $Z^M_A$ 
by left and right multiplication, respectively. In line with our identification
of $Z^M_A$ as the winding numbers $\d_A X^M$ of the membrane, one may
thus associate $SL(3)_{L}$ and $SL(3)_R$ with 
the modular groups of the world-volume 
and target space
3-torii. The remaining $SL(2)$ $U$-duality group,
acting by fractional linear transformations on the complex modulus 
$\tau=C+i\frac{V}{l_M^3}$ (with $C\equiv C_{123}$) must reside in the 
remaining non-linearly acting $SL(3)_{NL}$. A further decomposition
\be
SL(3)_{NL}\supset SL(2)_{\tau}\times\Real^+_\nu
\ee
allows the remaining $\Real^+_\nu$ factor to be identified with
the volume of the world-volume 3-torus. In the classical membrane
theory, the membrane volume 
does not decouple because, in contrast to strings, the
world-volume theory is not Weyl invariant.

\begin{figure}
{\small
\hspace{5.5cm}
\begin{picture}(190,60)
\thicklines
\multiput(0,0)(30,0){5}{\circle{8}}
\put(0,-12){\makebox(0,0){1}}\put(0,-25){\makebox(0,0){$\alpha_{2}^{3}$}}
\put(30,-12){\makebox(0,0){3}}\put(30,-25){\makebox(0,0){$\alpha_{1}^{2}$}}
\put(60,-12){\makebox(0,0){4}}\put(60,-25){\makebox(0,0){$\alpha_{1}^{\1d}$}}
\put(90,-12){\makebox(0,0){5}}\put(90,-25){\makebox(0,0){$\alpha^{\1d}_{\2d}$}}
\put(120,-12){\makebox(0,0){6}}\put(120,-25){\makebox(0,0){$\alpha^{\2d}_{\3d}$}}
\multiput(4,0)(30,0){4}{\line(1,0){22}}
\put(60,4){\line(0,1){22}}
\put(60,30){\circle{8}}
\put(58,57){\!$\times$}
\put(75,30){\makebox(0,0){$\beta_0$}}
\put(48,30){\makebox(0,0){2}}\put(75,60){\makebox(0,0){$-\omega$}}
\put(61.1,41){$\!\vdots$}
\end{picture}}
\vskip 1cm
{\footnotesize
$$\begin{array}{cc}
\alpha_{1}^{2}= &      (0,0,1,0,0,0)\\[.7mm] 
\alpha_{2}^{3}= &      (1,0,0,0,0,0)\\[.7mm] 
\alpha_{1}^{3}= &      (1,0,1,0,0,0)\\[.7mm] 
\alpha_{\2d}^{\1d}= &      (0,0,0,0,1,0)\\[.7mm] 
\alpha_{\3d}^{\2d}= &      (0,0,0,0,0,1)\\[.7mm] 
\alpha_{\3d}^{\1d}= &      (0,0,0,0,1,1)\\[.7mm] 
\end{array}$$
$$\begin{array}{cc}
\alpha_{\1d}^{1}= &      (0,0,0,1,0,0)\\[.7mm] 
\alpha_{\1d}^{2}= &      (0,0,0,1,1,0)\\[.7mm]
\alpha_{\1d}^{3}= &   (0,0,0,1,1,1)\\[.7mm] 
\alpha_{\2d}^{1}= &      (0,0,1,1,0,0)\\[.7mm] 
\alpha_{\2d}^{2}= &   (0,0,1,1,1,0)\\[.7mm] 
\alpha_{\2d}^{3}= &   (0,0,1,1,1,1)\\[.7mm] 
\alpha_{\3d}^{1}= &   (1,0,1,1,0,0)\\[.7mm] 
\alpha_{\3d}^{2}= &   (1,0,1,1,1,0)\\[.7mm] 
\alpha_{\3d}^{3}= &   (1,0,1,1,1,1)\\[.7mm] 
\end{array}$$
$$\begin{array}{cc@{\hspace{7mm}}c}
\beta_0= &   (0,1,0,0,0,0)  &   \gamma_0=(1,1,2,3,2,1)\\[.7mm]
\beta_{\1d}^{1}= &   (0,1,0,1,0,0)  &   \gamma_{1}^{\1d}=(1,1,2,2,2,1)\\[.7mm]
\beta_{\1d}^{2}= &   (0,1,0,1,1,0)  &   \gamma_{1}^{\2d}=(1,1,2,2,1,1)\\[.7mm]
\beta_{\1d}^{3}= &   (0,1,0,1,1,1)  &   \gamma_{1}^{\3d}=(1,1,2,2,1,0) \\[.7mm]
\beta_{\2d}^{1}= &   (0,1,1,1,0,0)  &   \gamma_{2}^{\1d}=(1,1,1,2,2,1)\\[.7mm]
\beta_{\2d}^{2}= &   (0,1,1,1,1,0)  &   \gamma_{2}^{\2d}=(1,1,1,2,1,1) \\[.7mm]
\beta_{\2d}^{3}= &   (0,1,1,1,1,1)  &   \gamma_{2}^{\3d}=(1,1,1,2,1,0) \\[.7mm]
\beta_{\3d}^{1}= &   (1,1,1,1,0,0)  &   \gamma_{3}^{\1d}=(0,1,1,2,2,1)\\[.7mm]
\beta_{\3d}^{2}= &   (1,1,1,1,1,0)  &   \gamma_{3}^{\2d}=(0,1,1,2,1,1) \\[.7mm]
\beta_{\3d}^{3}= &   (1,1,1,1,1,1)  &   \gamma_{3}^{\3d}=(0,1,1,2,1,0) \\[.7mm]
\end{array}$$
$$\begin{array}{cc}
\omega = &   (1,2,2,3,2,1)  
\end{array}$$}
\vspace{-.8cm}
\caption{Dynkin diagram and positive roots for $E_6$}
\label{E6}
\end{figure}

One may wonder if this choice of parameterization is ambiguous.
In particular, one may have considered identifying the world volume
modular group with the non-linearly realized factor $SL(3)_{NL}$.
However, thanks to the triality symmetry 
of the extended $E_6$ Dynkin diagram in Figure~\ref{E6}, 
one can show
by using an appropriate intertwiner 
that any two of the $SL(3)_{L,R,NL}$ groups can be made to act linearly.
Identifying the world-volume
modular group with the linearly realized $SL(3)_L$ is 
convenient for our purposes, since we wish to integrate over 
the moduli space of world-volume metrics $SO(3)\backslash SL(3)$.
Moreover, the combination of linearly acting world-volume and
target-space modular groups is  necessary for our membrane 
interpretation of the exceptional theta series.

To determine the decomposition of $SL(3)_{NL}$ into world-volume
and $U$-duality parts, we recall (as observed in
Section~\ref{minrep}) that the target three-form 
couples through the exponential of the generator $e_{-\beta_0}$.
Therefore we associate the non-linear $SL(2)$ $U$-duality group
with that generated by $\{E_{\beta_0},H_{\beta_0},E_{-\beta_0}\}$.
The maximal commutant to the $SL(2)\subset SL(3)_{NL}$ is the
$\Real^+$ group generated by
\be
H_\nu=2H_{\omega}-H_{\beta_0} = -3 y\p_y -3 x_0 \p_{x_0} -2 Z\cdot
\p_Z -12
\, .
\ee
{}From the expressions for the $E_6$ generators in
Figure~\ref{E6_rep} we read off the scaling weights of the
various quantum numbers with respect to target and world volumes:
\be
\begin{array}{|c|ccc|}
\hline
\mbox{ quantum numbers }&\ y \ &\ x_0 \ &\ Z_A^M\ \\
\hline
\mbox{ target-space }&-1\ &1\ &0\ \ \\
\mbox{ world-volume }&-3\ &-3\ &-2\ \ \\
\hline
\end{array}
\ee
Therefore we choose the following couplings
\be
y\rightarrow \nu V^{1/3}y\, ,\qquad
x_0\rightarrow \nu V^{-1/3}x_0\, ,\qquad
Z\rightarrow \nu^{2/3} e^{-1} Z E \, ,
\ee
where the $3\times 3$ matrices
\be
\gamma\equiv\nu^{2/3}\ e e^\rt \, ,\qquad
G\equiv V^{2/3}\ E E^\rt \, ,
\label{metrics}
\ee
are the world-volume and target-space metrics, respectively.
Note that although relative scalings of the two volumes
amongst the variables $(y,x_0,Z_A^M)$ are fixed, we will justify
the overall ones by the results. We may now state our 
$U$-duality and world-volume modular 
invariant membrane winding formula
\be
\theta_{E_6}\!(\gamma;G,C)=V\ \nu^2\!\!\!\!
\sum_{\stackrel{(y,x_0,Z)}{\sss \in \Zint^{11}\setminus\{0\}}}
{\ss \mu(y,x_0,Z)}\ \frac{e^{
-\frac{2\pi}{l_M^3}\frac{\sqrt{\det(ZGZ^\rt 
+\gamma
|x_0+\tau y|^2)
}}
{|x_0+\tau y|^2}
-2\pi i\det Z
\big(
\frac{x_0+C y}{y|x_0+\tau y|^2}
\big)
}}
{\ss\sqrt{\det(ZGZ^\rt 
+\gamma
|x_0+\tau y|^2)
} }\
+ \ \mbox{\small degen.} \label{theta}
\ee
Salient features of this result include
\begin{itemize}
\item Although the variables $Z_A^M$ of the minimal
representation in Figure~\ref{E6_rep} are real-valued, once
integrated against the distribution $\delta$ in~\eqn{formula},
they are restricted to integers. (This is also the origin of
the overall factors $2\pi$ in the exponent.) Hence their natural
interpretation as winding numbers.
\item The real exponent is a Born-Infeld membrane action. 
It interpolates between Nambu-Goto (large membrane volume
$V\gg \nu$) and Polyakov-like ($V\ll \nu$) actions.
\item The membrane tension appears correctly as $1/l_M^3$
while the overall factor of the target space volume $V$  matches
correctly that of the bulk term in~\eqn{instantons}.   
\item The subleading degenerate terms and summation measure $\mu(y,x_0,Z)$ 
are known~\cite{KazPol} and described in Appendix~\ref{measdeg}. 
The latter is derived from a $p$-adic analog of the
spherical vector. It is a complicated number theoretic function
representing the quantum degeneracies of winding states and was
inaccessible to previous semi-classical
approaches~\cite{Pioline:2001jn}.
\item The $SL(2)$ modulus $\tau=C+i\frac{V}{l_M^3}$ coupling 
to the fluxes $(y,x_0)$ may be written covariantly as 
$|H_{MNR}+C_{MNR}H|^2+|H|^2$ where $x_0\leftrightarrow
H_{MNR}=dB_{MNR}$ and $y\leftrightarrow H\equiv dB$. The
interpretation of $x_0$ as a target space 3-form is justified 
both by its coupling to  the Chern-Simons 3-form and generalizations
to higher torii discussed in Section~\ref{highertorii}.  
\end{itemize}
The simplest check of our proposal is whether it is a zero mode
of the Laplacian on the $U$-duality moduli space, as required by
supersymmetry. This is the topic of the next Section.

\subsection{$E_6$ Casimirs}
\label{cassies}

The desired $R^4$ amplitude~\eqn{fr4} is a zero mode
of the Laplacian and invariant cubic operator of the
$SL(3)\times SL(2)$ $U$-duality group. This is in fact separately true 
for the geometric target space $SL(3)$ subgroup (and also the Laplacian of the 
non-linear $SL(2)$ factor with which we deal later). We can easily
verify this property of our $E_6$ based $R^4$ amplitude as a simple
consistency check: By virtue of the formula~\eqn{formula}, relations
valid for the enveloping algebra of the representation $\rho$ apply
also to the corresponding (differential) operators acting on moduli
$g$.
We therefore examine the quadratic and cubic Casimirs of the
subgroups $SL(3)_{L,R,NL}$,
\bea
{\cal C}_2&=&\:\frac{1}{3!}\ \sum_{\alpha\in\Delta^+}\Big[H_\alpha^{\
\! 2}
-6\, E_\alpha
E_{-\alpha}\Big]_{\rm\sss Weyl}\, ,\\
{\cal C}_3&=&\frac12 \prod_{\alpha\in\Delta^+}\!\widetilde H_\alpha
+\ \frac{9}{2}\ \Big[\sum_{\alpha\in\Delta_+}\widetilde H_\alpha E_\alpha
E_{-\alpha}
-3\!\!\!
\sum_{\begin{array}{c}\sss\alpha,\beta,\gamma\in\Delta^+\\[-3mm]
\sss\alpha+\beta=\gamma
\end{array}}\!\!\!
\Big(E_{\alpha}E_{\beta}E_{-\gamma}
+E_{-\alpha}E_{-\beta}E_{\gamma}\Big)
\Big]_{\rm\sss Weyl} .\nn\\
\label{casimirs}
\eea
Here $\alpha_{1,2}$ are simple roots and
$\Delta^+=\{\alpha_{1,2},\alpha_{1}+
\alpha_2\}$ is the positive root lattice (see~\eqn{sl3s}).
These compact expressions are ``classical'', the square brackets
denote Weyl ordering averaging each term over all distinct orderings.
Also, $\widetilde H_{\alpha_{1}}\equiv 
H_{\alpha_{1}}+H_{\alpha_1+\alpha_2}$,
$\widetilde H_{\alpha_{2}}\equiv 
-H_{\alpha_{2}}-H_{\alpha_1+\alpha_2}$
and $\widetilde H_{\alpha_1+\alpha_2}\equiv H_{\alpha_1}-H_{\alpha_2}$.
Inserting the explicit expressions for the minimal representation
$E_6$ generators (see Figure~\ref{E6_rep}) yields invariant differential
operators $\Delta^{(2,3)}_{SL(3)_{R,L,NL}}$ subject to particular
relations
\be
\Delta^{(2)}_{SL(3)_L} = \Delta^{(2)}_{SL(3)_R} =
\Delta^{(2)}_{SL(3)_{NL}}
\, , \qquad
\Delta^{(3)}_{SL(3)_L} = \Delta^{(3)}_{SL(3)_R} =
\Delta^{(3)}_{SL(3)_{NL}}\, . \label{cassanovas}
\ee
As explained above in~\eqref{dso}, 
supersymmetry requires that the invariant target space 
operators $\Delta^{(2)}_{SL(3)_R}=0= \Delta^{(3)}_{SL(3)_R}$.
However, upon integrating over the $SL(3)_L$
fundamental domain of the world-volume torus shape moduli, the
operators
$\Delta^{(2,3)}_{SL(3)_L}$ no longer act and must therefore return
zero. Hence~\eqn{cassanovas}, in turn, implies vanishing
of the target space invariants.
As in the 1-loop string computation~\cite{Dixon:1990pc}, infrared divergences
may lead to holomorphic anomalies, and a constant non-vanishing
right-hand side for~\eqref{dso}.  
The explicit world-volume moduli
integration is the subject of the following Section.

\section{Integrating over membrane world-volume moduli}
\label{wvintegrals}

An important difference between supermembranes and superstrings
is the absence of a classical Weyl symmetry. An integral over all 
membrane volumes is, in general, divergent and must be appropriately
regulated. On the other hand, integrations over $SL(3)$ shape moduli
of a world-volume 3-torus are better defined. This part of our
calculation is completely analogous to its stringy counterpart: The
summation over windings is decomposed into $SL(3,\Zint)$ orbits 
and the integration over shape moduli can be performed by
unfolding their fundamental domain. 

An additional puzzle, at this stage, is the r\^ole of the flux-like
quantum numbers~$(y,x_0)$. Our solution is to relate this difficulty
to regulating the volume integral: Replacing the volume integral
by one over 
additional compact moduli corresponding to shifts of the fluxes
at the same time integrates out these additional quantum numbers
while projecting out the dependence on the volume modulus. This
procedure is a general technique in the theory of automorphic forms
known as a constant term computation\footnote{An excellent review
and useful results for $SL(3)$ can be found in~\cite{Miller}.}.
We begin the computation by integrating over shape moduli.

\subsection{Integration over $SL(3)$ shape moduli}

The next step of our investigation of the conjecture that 
\be
Z_{\rm mem}(\gamma_{AB}; G_{MN},C_{MNP})=\theta_{E_6}(\gamma_{AB};
G_{MN},C_{MNP})\, ,
\ee
is to evaluate the modular integral over 
the $SO(3)\backslash SL(3)$ shape moduli of the world-volume 3-torus. 
Let us begin with a general description of the $SL(3)_L$ modular integral: 

\subsubsection{Modular integral and orbit decomposition}

The theta series summand at an arbitrary point
of the moduli space $SO(3)\backslash SL(3)$, with Iwasawa gauge coset representative 
\be
e^{-1}=
\begin{pmatrix} 1/L && \\ &\sqrt{L/T_2} & \\ && \sqrt{L\,  T_2}
\end{pmatrix} \cdot
\begin{pmatrix} 1 & A_1 & A_2 \\ & 1 & T_1 \\ & & 1
\end{pmatrix} \, ,
\label{Iwasawa}
\ee 
is given in~\eqn{theta} depending on moduli $e$ through~\eqn{metrics}. 
This summand was obtained by acting on the spherical 
vector with the linear $SL(3)_L$ representation
$\rho_L(e)$ acting by left multiplication on the windings $Z^M_A$.
One is therefore left to compute
\be
\label{modint}
\theta_{Sl_3\times Sl_3}=\int_{\cal F} de\ 
\theta_{E_6}(y,x_0,e^{-1}Z)\, .
\ee
The integration is over the fundamental domain 
${\cal F}=SO(3)\backslash SL(3)_L/SL(3,\Zint)$ of the moduli space
of unit-volume constant metrics on $T^3$, with 
invariant measure 
\be
de=\frac{d^2T}{T_2^2}\ d^2\!A\ \frac{dL}{L^4}\, .
\ee
We lose no generality 
evaluating all other (target space) moduli at the origin.
We stress again, that by construction $\theta_{Sl_3\times Sl_3}$
is an automorphic form of the $U$-duality group $SL(3,\Zint)\times
SL(2,\Zint)\subset SL(3,\Zint)\times
SL(3,\Zint)$.

Modular integrals of this type can be computed by the general method 
of orbits: one restricts the summation on integers in $\theta_{E_6}$
to one representative in each orbit of the linear $SL(3,\Zint)_L$ action, 
and at the same time enlarges the integration domain to the image of the
fundamental domain under the $SL(3,\Zint)$ orbit generators.
Since $Z$ transforms by $SL(3,\Zint)$ left-multiplication,
orbits are labeled by the rank of the $3\times 3$ matrix $Z$.
The non-degenerate orbits have
${\rm rank}(Z)=3$ and the fundamental domain can be enlarged to
the full $SO(3)\backslash SL(3)$ moduli space. At the other end of the rank spectrum, 
the ${\rm rank}(Z)=0$ orbit contains the single element
$Z=0$. The shape integral then yields only a factor
of the volume of the fundamental domain ${\cal F}$.
After integration over the volume factor $\Real^+_\nu$, these two orbits
will correspond to the Eisenstein series $E^{SL(2)}_{{\bf 2},1}$
in~\eqn{fr4}, hence to membrane instantons.
The ${\rm rank}(Z)=1,2$ orbits require a detailed understanding of
the fundamental domain ${\cal F}$. They correspond to toroidal,
supergravity Kaluza--Klein excitations with amplitude $E^{SL(3)}_{{\bf
3},3/2}$. We deal with these terms first for which we
are able to present a rigorous computation:

\subsubsection{Rank 1 and 2 winding modes}

To performing the integral over shape moduli it is convenient to 
rewrite the $E_6$ spherical vector in the integral
representation
\be
f_{E_6} = \frac{1}{\sqrt{\pi}}\int_0^\infty \frac{dt}{t^{1/2}}
\exp\left( - \frac{1}{4t (y^2+x_0^2)^2} - t \det[ ZZ^\rt  + (y^2+x_0^2) 
\mathbb{I}] - i \frac{x_0 \det(Z)}{y (y^2+x_0^2)} \right)\, .
\label{intrep}
\ee
Since $SL(3)_L$ acts on the matrix $Z$ by left multiplication,
$Z \to e^{-1} Z$, it leaves the phase invariant. We may further
set $x_0=0$ as it can be reinstated by an $SO(2)$ rotation in the
$(y,x_0)$ plane. 

We must now compute $\int_{{\cal F}_{{\rm rk= 1,2}}}\!\!de\ f_{E_6}$
with respect to unfolded fundamental domains ${\cal F}_{{\rm rk=
1,2}}$.
The $SL(3)$ fundamental domain ${\cal F}$ 
is known~\cite{Minkowski,Grenier}, 
a complete description is given in Appendix~\ref{FDSL3},
see in particular equations~\eqn{FD1} and~\eqn{FD2}. Its construction
is in terms of a height function given by the maximal abelian torus
coordinate $L^3$ in~\eqn{Iwasawa} along with the actions of an
(overcomplete) set of $SL(3,\Zint)$ generators ${\rm S}_{1,\ldots,5}$, 
${\rm T}_{1,2}$ and ${\rm U}_{1,2}$ given in~(\ref{STU1},\ref{STU2}). 

\vspace{.2cm}
\noindent
{\bf Rank 1 (twice degenerate) orbit:} In this case the winding matrix $Z$ 
may be rotated  by an $SL(3,\Zint)$ transformation from the right into
a single row
\be
Z=\begin{pmatrix}\  p_1 & p_2 & p_3  \\ & & & \\ & & \\
\end{pmatrix} 
\, .
\label{Z1}
\ee
The range of the linear mapping $Z$ is generated 
by a single $SL(3)_L$ left-invariant vector transforming
as a projective vector under the right $SL(3)_R$ action.
The matrix~\eqn{Z1} is
left invariant 
with respect to $SL(3,\Zint)$ elements of the form
\be
\begin{pmatrix}
\ 1& \ *\  & *\  \\
 & * & *\  \\
 & * & *\ 
\end{pmatrix}\, ,
\ee
spanned by generators ${\rm S}_2$, ${\rm T}_{1,2,3}$ and ${\rm U}_2$. The conditions
from the remaining generators may be unfolded yielding the integration
range
\be
{\cal F}_{{\rm rk=1}}=
\Big\{
0\leq T_1\leq \frac12\, ,\;
T_1^2+T_2^2\geq1\, ,\;
-\frac12\leq A_{1,2}\leq\frac12\, ,\;
0\leq L<\infty
\Big\}\, .
\ee
We find
\bea
\int_{{\cal F}_{{\rm rk= 1}}}\!\!de\ f_{E_6}&=&\frac{1}{\sqrt{\pi}}
\int_{{\cal F}_{{\rm rk= 1}}}\!\!de\int\frac{dt}{t^{1/2}}\ 
\exp\Big(-\frac{1}{4y^4t}-ty^4\Big[y^2+\frac{p_1^2+p_2^2+p_3^2}{L^2}
\Big]
\Big)\nn\\
&=&
\frac{1}{(p_1^2+p_2^2+p_3^2)^{3/2}}\ 
\frac{\pi}{3|y|}\ K_1(|y|)\, .
\eea
The first factor is easily recognized as an Eisenstein series
of $SL(3)_R$ in the minimal representation, with index $3/2$.
On the other hand, reinstating the $x_0$ dependence dependence 
in the second term, $|y|\rightarrow\sqrt{y^2+x_0^2}$, yields
the spherical vector ~\eqn{fourminsph} of the Eisenstein series
of $SL(3)_{NL}$ in the minimal representation, with index 0.
Both of these can be identified with the general Eisenstein 
series \eqref{EisSl3} with $(\lambda_{32},\lambda_{21})$ equal 
to $(-2,1)_R$ and $(1,1)_{NL}$, respectively. 
Hence, up to an overall normalization\footnote{The rank~1 and~2
normalizations are computable, but unimportant since we have no such
control over the rank~3 computation.  Note also, there is in principle
a six-fold ambiguity in the above identification, which
is irrelevant due to the Selberg functional relations
discussed in Appendix~\ref{sl3-rep}.} 
we obtain, in the notation of Appendix~\ref{eis_app},
\be
\int_{{\cal F}}de\ \theta_{E_6}^{\rm rk=1}
=E(g_{R};-2,1)\ E(g_{NL};1,1)
\, .
\ee
The $SL(3)_R$ moduli are $g_R=EE^\rt $, the unit, target space metric. We
postpone dealing with the $SL(3)_{NL}$ moduli $g_{NL}$
to our discussion of the membrane
world-volume integration in Section 4.2. 

\vspace{.2cm}
\noindent
{\bf Rank 2 (singly degenerate) orbit:}
The winding matrix $Z$ 
may be rotated  by an $SL(3,\Zint)$ transformation into
two rows,
\be
Z=\begin{pmatrix} a_1 & p_1 & p_2 \\ & a_2 & p_3 \\ & & \\
\end{pmatrix}
\equiv\begin{pmatrix}  & \vec v &  \\ & \vec w &  \\ & & \\
\end{pmatrix}\, .
\ee
This configuration is invariant under the left $SL(3,\Zint)$ action
of elements of the form
\be
\begin{pmatrix}
\ 1&   & *\  \\
 & \ 1\  & *\  \\
 &  & *\ 
\end{pmatrix}\, .
\ee
These are spanned by generators $T_{2,3}$. All other generators may be
unfolded yielding the integration range
\be
{\cal F}_{{\rm rk =2}}=
\Big\{
-\frac12\leq T_1\, ,\; A_2\leq \frac12\, ,\;
-\infty<A_1<\infty\, ,\;
0\leq T_2,L<\infty
\Big\}\, .
\ee
We find
\bea
\int_{{\cal F}_{{\rm rk= 2}}}\!\!de\ f_{E_6}&=&\frac{1}{\sqrt{\pi}}
\int_{{\cal F}_{{\rm rk= 2}}}\!\!de\int\frac{dt}{t^{1/2}}\ 
e^{-\frac{1}{4y^4t}-ty^2
\Big[y^4+
y^2\Big(
\frac{L\vec v^2}{T_2}+\frac{(\vec w+A_1\vec v^2)}{L^2}
\Big)
+\frac{|\vec v\times \vec w|^2}{LT_2}
\Big]
}\nn\\
&=&
\frac{1}{|\vec v \times\vec w|^3}\ 
\frac{1}{y^2}\int_0^\infty\frac{dt\ dT_2 \ dL}
{t \ T_2^2\ L^3}
e^
{
-\frac{1}{4ty^4}-ty^2\left( y^2+ \frac{1}{L^2}  \right)
\left( y^2+ \frac{L}{T_2}  \right)  
}\, .\nn\\
\eea
Here the integrals over compact circles $T_1$ and $A_2$ are trivial
and we performed the Gaussian integral over $A_1$ explicitly. 
We now change variables 
\be
r_1=1/L^2\, ,\qquad r_2=L/T_2\, ,\label{vc}
\ee
and rescale $r_i\to r_i  y^2$, $t\to t/y^5$, obtaining
\bea
\int_{{\cal F}_{{\rm rk= 2}}}\!\!de\ f_{E_6}&=&
\frac{y^3 }{2|\vec v \times\vec w|^3}\ 
\int_0^\infty\frac{dt}{t}
\sqrt{r_1} dr_1 dr_2~
e^
{
-\frac{y}{4t}-ty( 1 + r_1) (1+r_2)
}\, .\nn\\
\eea
The integral over $r_2$ is of Gamma function type. Carrying out
the Bessel-type
integration with respect to $t$ 
and changing variable to $r_1=u^2-1$, we find
\bea
\int_{{\cal F}_{{\rm rk= 2}}}\!\!de\ f_{E_6}&=&   
\frac{4y^2}{|\vec v \times\vec w|^3}\ 
\int_{1}^{\infty} K_1(u y) \sqrt{u^2-1} ~du\\
&&=
\frac{2\pi e^{-y}}{|\vec v \times\vec w|^3} \nn
\eea
Comparing again to~\eqref{EisSl3} and~\eqref{fourminsph},
we recognize the product of $SL(3)_{NL}$
and $SL(3)_R$ continuous representations with parameters
$(\lambda_{32},\lambda_{21})=(1,1)$, $(1,-2)$, respectively.
Hence the rank~2 result is a product of corresponding
$SL(3)$ Eisenstein series. However, as explained in 
Appendix~\ref{constant_term_app}, minimal parabolic Eisenstein 
series obey Selberg relations, which amount to invariance\footnote{The
classical example is the $SL(2)$ relation $E^{SL(2)}_s\propto
E^{SL(2)}_{1-s}$. Equality holds for appropriate normalization by
a function of $s$, see~\eqn{Selberg}.} 
under (Weyl group) permutations of the labels
$\lambda_{1},\lambda_{2},\lambda_{3}$. These are generated by
reflections about radial lines $\lambda_{21}=\lambda_{32}$, 
$2\lambda_{21}=-\lambda_{32}$ and $\lambda_{21}=-2\lambda_{32}$
in the $(\lambda_{32},\lambda_{21})$-plane depicted in Figure~\ref{sl3cas}. 
In particular, this implies, up to normalization, equality of the 
rank~2 and rank~1 winding sums. Hence
\be
\int_{{\cal F}}de\ \Big(\theta_{E_6}^{\rm rk=1}+\theta_{E_6}^{\rm
rk=2}\Big)
=E(g_{NL};-2,1)\ 
E(g_{R};-2,1)\, .
\label{rk12}
\ee
Observe from Figure~\ref{sl3cas} that the point $(-2,1)$ corresponds to
vanishing quadratic and cubic Casimirs as predicted in
Section~\ref{cassies}.

\subsection{Integration over membrane volume -- degenerate orbits}

Before dealing with rank~3 and~0 winding sums, we study 
the rank~2 and~1 results and learn how to handle the membrane
world volume integral. The lack of conformal (Weyl) invariance of the classical
supermembrane theory has been a key stumbling block, intimately
related to its gapless, continuous spectrum. This is precisely the
sector where we expect to find new physics.

We start with an Iwasawa decomposition of $SL(3)_{NL}$ moduli 
refined to exhibit the $SL(2)$ $U$-duality subgroup:
\be
g_{NL}=
\begin{pmatrix}
\frac1{\nu^{2/3}}&&\\&\nu^{1/3}&\\&&\nu^{1/3}
\end{pmatrix}
\begin{pmatrix}
1&&\\&\frac{1}{\sqrt{\tau_2}}&\\&&\sqrt{\tau_2}
\end{pmatrix}
\begin{pmatrix}
1&&\\&\ 1\ &\tau_1\\&&1
\end{pmatrix}
\begin{pmatrix}
1&n_1&n_2\\&1&\\&&1
\end{pmatrix}\, .
\ee
The membrane world volume modulus $\nu$ and target space moduli
$(\tau_1,\tau_2)$ were already introduced in Section~\ref{paramsect}.
There are, however, two additional possible moduli $(n_1,n_2)$.
Automorphy implies that these are compact unit interval valued
variables. In contrast, the world volume modulus is non-compact
taking any real positive value. In a (semi)classical supermembrane
setting, the moduli $(n_1,n_2)$
have no particular meaning and should be set to zero;
one would further have to integrate over all world volumes
$\nu\in\Real^+$, which is an ill-defined non-compact integral.
In particular this integral can be performed explicitly 
for the rank~1 and~2 winding sums~\eqn{rk12} and seen to be divergent
(see Appendix~\ref{wv_app} for a sample computation).

Instead we propose exchanging the world-volume modulus $\nu$ for the
compact moduli $(n_1,n_2)$. An integral over these moduli is well
defined, and known as a constant term computation in the 
mathematical literature. Physically, we
could then view $(y,x_0)$ as auxiliary quantum numbers, necessary for
$U$-duality based on a hidden exceptional symmetry group. 
They are ``integrated out'' by performing the $(n_1,n_2)$ integrals.
Indeed, $(y,x_0)$ and $(n_1,n_2)$ are world volume
canonical conjugates. Essentially we are adding auxiliary world volume
fields to make the exceptional symmetry manifest, and in turn
integrating them out. The constant term result is then invariant under 
a subgroup $SL(2)$, the Levi part of the parabolic group $P_2$ with
unipotent radical spanned by $(n_1,n_2)$.

Placing faith in our proposal we must now compute $\int_0^1 dn_1 dn_2
E(g_{NL};-2,1)$. Physicists can easily perform
this computation by using the sum representation of the Eisenstein 
series and Poisson resumming, {\it i.e.} a small radius expansion
in one direction~\cite{Obers:1999um}. A general mathematical machinery
involving $p$-adic integrations has been developed for these
computations
by Langlands~\cite{Langlands} (again a useful account is given
in~\cite{Miller} and for completeness these results are reproduced in
Appendix~\ref{constant_term_app}). Using~\eqn{constantterm}, 
the part of the result required
here is
\be
\int_0^1 dn_1 dn_2
E(g_{NL};1-2s,1)=\frac{\pi}{\zeta(3)}\frac{1}{s-3/2}+
{\rm analytic}\, .
\ee
In this normalization the leading behavior at $s=3/2$ is a simple
pole. Importantly it is $\nu$-independent! The subleading analytic
terms\footnote{In fact, since we have not studied the overall
normalization, one might argue that the leading contribution
in the product of Eisenstein series in~\eqn{rk12} is a double
pole with constant coefficient. The subleading single pole coefficients
then include the Dedekind eta in the $U$-duality modulus $\tau$. This
is in principle a feasible situation since this is the correct result
in the bulk wrapping sector. However, since there is anyway an implicit
additive renormalization of the $R^4$ conjecture~\eqn{fr4},
we prefer the above presentation. } 
depend both on the volume modulus $\nu$ and the log of the
Dedekind eta function of the $U$-duality modulus $\tau$. Nonetheless,
we claim that the correct prescription is to keep the coefficient of
the simple pole,
\be
\int d^2n \int_{\cal F} de\ \Big(\theta_{E_6}^{\rm rk=1}+\theta_{E_6}^{\rm
rk=2}\Big)
\propto E^{SL(3)}_{{\bf 3},3/2}(\widehat G)\, .
\ee
It remains only to derive the $SL(2)$ Eisenstein series part of the
the $R^4$ amplitude~\eqn{fr4}, from the rank 0 and rank 3 orbits.

\subsection{Membrane wrapping sum}
\label{Xmas}

The rank~0 and~3 contributions to the membrane winding summation
are independent of
the unit $SL(3)_R$ target space metric moduli and
correspond to the summation over membrane wrappings
in~\eqn{instantons}. Indeed the determinant of the winding matrix $Z$ 
counts toroidal wrappings and corresponds to $m^{[3]}$ there. 
The rank~0 contribution amounts to setting $Z=0$ and since no
unfolding is possible, simply
returns the volume of the fundamental domain ${\cal F}$.  
A {\it tour de force} calculation would track these degenerate
contributions by taking account also those of the original $E_6$ theta
series in~\eqn{theta} (described in Appendix~\ref{measdeg}). Here we
are rather less ambitious, however, since the final result is
guaranteed to be $U$-duality invariant, these terms are anyway fixed
by automorphy and we will not consider them further.
That leaves only the (most difficult) rank~3 terms:

Although we have gathered a great deal of information about the 
non-degenerate rank~3 term, unlike the lower rank contributions, we are
unable to compute it exactly. We are able to (i) calculate an approximate
expression for the $SL(3)_{NL}$ spherical vector appearing after
performing membrane shape moduli integrals; (ii) identify the
underlying (novel) $SL(3)_{NL}$ 
representation on which this spherical vector is based;
(iii) compute the intertwiner between this representation of
$SL(3)_{NL}$ and the continuous series representation of $SL(3)$;
(iv) compute the action of this intertwiner on the approximate
spherical vector although this does not yield an unambiguous
identification of the $SL(3)_{NL}$ automorphic form at hand.
Points (i)-(iv) are presented chronologically in what follows.
As evidence that upon integration over compact moduli $(n_1,n_2)$
the result is the required $SL(2)$ Eisenstein series, we employ the
intertwiner of point (iii) to study the constant computation for the
standard induced representation and show that the leading contribution is the
correct one.

\subsubsection{Rank~3 winding modes}

An element $Z$ of the non-degenerate orbit can be
rotated by an $SL(3,\Zint)$ matrix into
\be
Z=\begin{pmatrix} a_1 & p & q \\ & a_2 & r \\ & & a_3 \end{pmatrix}
\ee
No $SL(3,\Zint)$ generators leave this orbit representative
invariant, so the fundamental domain can be completely unfolded to
\be
{\cal F}_{{\rm rk =3}}=
\Big\{
-\infty<T_1,A_1,A_2<\infty\, ,\;
0\leq T_2,L<\infty
\Big\}\, .
\ee
We again employ the integral representation of the spherical
vector~\eqn{intrep} and the change of world volume shape
variables~\eqn{vc}. 
The integral with respect to $A_{2}$ and $T_1$ is Gaussian.
The subsequent integral over $A_1$ leads to
\be
\begin{split}
&\int\frac{dt dr_1 dr_2}{t^{3/2}a_2a_3^2 r_2^{1/2} y^3}  \exp\left[-\frac{1}{4ty^4}-t\frac{(r_1a_1^2+y^2)(r_2a_2^2+y^2)(a_3^2+r_1r_2y^2)}{2r_1r_2}\right]\\
&\qquad \times K_0\left( \frac{t(r_1a_1^2+y^2)(r_2a_2^2+y^2)(a_3^2+r_1r_2y^2)}{2r_1r_2}\right) .
\end{split}
\ee
In the limit where the argument of $K_0$ is very large, this reduces to
\be
\int \frac{r_1 dr_1 dr_2 dt }{t^{2}}
\frac{\exp\left( - \frac{1}{4t y^4} 
- t \frac{(a_1^2 r_1+y^2)(a_2^2 r_2+y^2)(a_3^2+y^2 r_1 r_2)}{r_1 r_2}  \right)}
{r_1^{1/2} a_2 a_3^2 y^3 
\sqrt{(a_1^2 r_1+y^2)(a_2^2 r_2+y^2)(a_3^2+y^2 r_1 r_2)}}\, .
\ee
The integral over $r_1,r_2,t$ can be performed 
in the saddle point approximation, and finally gives
\be
f_{SL(3)}= \int_{{\cal F}_{\rm rk=3}} de f_{E_6}
\sim \frac
{\exp\left[ -\frac{(y^2+x_0^2+x_1^2)^{3/2}}{y^2+x_0^2} 
- i \frac{x_0 x_1^3}{y (y^2+x_0^2)} \right]}
{x_1^5(y^2+x_0^2+x_1^2)^{1/4}}
\label{E6A2sph}
\ee
where $x_1^3 \equiv a_1 a_2 a_3=\det Z$ is the wrapping number.
 This saddle point 
result for the spherical vector becomes exact in the limit where
$(y,x_0,x_1)$ are scaled to infinity at the same rate. The fact
that the result depends on the determinant of the matrix $Z$ is guaranteed by 
$SL(3)_L$-invariance, and implies that the result is an $SL(3)_R$
singlet, as it should if it is to reproduce the $SL(2)$ part
in~\eqref{fr4} after integration over the volume factor.
The representation under $SL(3)_{NL}$ is however more tricky
to identify.

\subsubsection{Representation of the non-degenerate orbit under $SL(3)_{NL}$}

Beginning with the $E_6$ minimal representation in Figure~\ref{E6_rep}, 
we can obtain
the representation of $SL(3)_{NL}$ by studying the action of the generators
$(E_{\pm\omega},E_{\pm\beta_0},E_{\pm\gamma_0},H_{\beta_0},H_{\gamma_0})$
restricted to functions $\varphi(y,x_0,z\equiv x_1^3=\det(Z))$. We find
\be
\begin{array}{lcll}
E_{\b_0}=y \p_{x_0}\, ,&\qquad&
E_{-\b_0}=&-x_0 \p_y - \frac{iz}{y^2}\, ,\\[2mm]
E_{\g_0}=i x_0\, ,&&
E_{-\g_0}=&-i(6+x_0\p_{x_0}+y\p_y+3z\p_z)\p_{x_0}\\
&&&-y(6+z^2\p_z^2+6z\p_z)\p_z\, ,\\[2mm]
E_{\om}=i y\, ,&&
E_{-\om}=&-i(6+x_0\p_{x_0}+y\p_y+3z\p_z)\p_y\\&&&
         +x(6+z^2\p_z^2+6z\p_z)\p_z\\&&&
         -\frac{3i}{y}\ (2+z^2\p_z^2+4z\p_z)
         +\frac1{y^2}\ z\p_y\, ,\\[2mm]
H_{\b_0}=x_0\p_{x_0}-y\p_y\,
,&&H_{\gamma_0}=&-2x_0\p_{x_0}-y\p_y-3z\p_z-6\, .
\end{array}
\label{sl3x1}
\ee
In Section~\ref{cassies} we argued that upon integrating out $SL(3)_L$
moduli, the remaining $SL(3)_{R,NL}$ Casimir invariants should vanish.
Indeed a straightforward computation yields
\be
\Delta^{(2)}_{SL(3)_{NL}} = 0= 
\Delta^{(3)}_{SL(3)_{NL}}\, .
\ee
The spherical vector for the representation~\eqref{sl3x1} may in principle
be computed by solving the partial differential equations associated
to the maximal compact subgroup $SO(3)$. We are not able to integrate
these equations exactly, however the leading result
in the limit where $y,x_0,x_1$ are scaled to infinity simultaneously
agrees with~\eqn{E6A2sph}.

\subsubsection{Intertwiner from cubic to induced representations}

We now identify the novel ``cubic''
$SL(3)_{NL}$ representation found
in~\eqref{sl3x1}. As we recall in Appendix~\ref{sl3-rep}, 
all continuous irreducible representations 
of $SL(3)$ can be obtained by induction from the minimal 
parabolic subgroup $P$ of
lower triangular matrices, with a character~\eqref{char}. 
These are natural candidates so long as the parameters
$(\lambda_{32},\lambda_{21})$ lie at the intersection of vanishing
quadratic and cubic Casimir loci depicted in Figure~\ref{sl3cas}:
\be
(\lambda_{32},\lambda_{21})\in \{
(-1,2),(1,1),(2,-1),(1,-2),(-1,-1),(-2,1) \}\ .
\ee
Note that these six solutions are related by action of the Weyl group,
hence the corresponding Eisenstein series by Selberg's relations~\eqref{selb3}.
We begin therefore with the representation~\eqn{sl3_cts} and search
for an intertwiner bringing it to the form~\eqn{sl3x1}. 
Let us now perform a
few changes of variables: 
We first Fourier transform over $v,w$, and write
$\p_w=ix_0, \p_v=iy$. The generator $-E_{\gamma}$ becomes
$y\p_0+\p_x\equiv E_{\beta_0}$. Similarly we identify $E_\beta$ with
$E_{\gamma_0}$.
In order to get rid of the $\p_x$ term, we
redefine 
\be
x \to x_1+x_0/y,\quad \p_x \to \p_1,\quad \p_0\to \p_0-\p_1/y, \quad
\p \to \p+x_0 \p_1/y^2\, .
\ee
The generator $H_{\beta_0}$ now becomes $-y\p+x_0\p_0+2 x_1 \p_1$. 
We eliminate the last term by further redefining 
\be
x_1=x_2/y^2\ ,\quad
\p_1 \to y^2 \p_2\ ,\quad \p_0 \to \p_0,\quad \p\to\p+2 x_2 \p_2/y
\ee 
The generator $E_{-\gamma_0}$
now reads $-x_0\p+x_2^2 \p_2/y^2$. We put $x_2=1/x_3$ so that
\be
E_{-\b_0}=-x_0\p-\frac{\p_3}{y^2}+(1-\lambda_{32})
\left( -\frac{x_0}{y}-\frac{1}{x_3 y^2} \right)
\ee
Only when $\lambda_{32}=1$, 
the singular term $1/(x_3 y^2)$ disappears,
so we may Fourier transform one last time over $x_3$
and write $\p_3=-i z$. This yields a one-parameter family of $SL(3)$
representations~\cite{leshouches} depending on $\lambda_{32}$.
Setting also $\lambda_{21}=1$, we obtain precisely the 
representation~\eqn{sl3x1}. We have thus identified the 
$SL(3)_{NL}$ representation arising by integrating the~$E_6$
theta series over the action of $SL(3)_L$ at a generic (rank~3)
point, with the Eisenstein series of $SL(3)_{NL}$ with
parameters $(\lambda_{32},\lambda_{21})=(1,1)$,
\be
\int_{{\cal F}}de\ \theta_{E_6}^{\rm rk=3}
\sim E(g_{NL};1,1)\, . 
\label{rk3}
\ee
Because the point $(1,1)$ 
lies at the intersection of two lines of single poles,
it is not clear however whether the right-hand side should be
understood as the residue, or whether some finite term should be kept -- 
we will return to this point shortly.

\subsubsection{Intertwining the spherical vector}

We begin with the known spherical vector~\eqn{spherical_A2} 
of the continuous representation
computed in Appendix~\ref{sl3-rep}. Because
$\lambda_{32}=\lambda_{21}=1$ in the intertwined $SL(3)_{NL}$
representation, we must study the limit $s=t=0$. 
We may however regard non-zero $s$ and $t$ as a regulator. 
In particular, representing the exact continuous representation
spherical vector as
\be
f_{SL(3)}=\frac{\pi^{s+t}}
{\Gamma(s)
\Gamma(t)}\,
\int_0^\infty 
\frac{dt_1dt_2}{t_1^{1+s}
t_2^{1+t}}\exp\left(
-\frac{\pi(1+x^2+(v+xw)^2)}{t_1}
-\frac{\pi(1+v^2+w^2)}{t_2}
\right) \, ,
\ee
and intertwining to the representation~\eqn{sl3x1} ({\it i.e.}
performing
the string of Fourier transforms and variable changes of the
preceding Section), in the saddle
approximation we recover 
the action appearing in the
exponent of~\eqn{E6A2sph}. The final $t_2$ integration yields an 
overall infinite factor, however. 
This situation is reminiscent
of the spherical vector for the maximal parabolic Eisenstein series
which can be obtained by a single Fourier transform over the variable
$w$. The rationale being that the continuous representation 
induced from the minimal parabolic with 
abelian character~\eqn{char}, 
is equivalent to that induced from the maximal parabolic with a
non-trivial representation in the $SL(2)$ Levi subgroup. The
possibility of attaching cusp forms correspondint to discrete
representations to the Levi factor, does
lead to independent Eisenstein series, however~\cite{Vahut}. 
Due to this type of subtlety, we cannot unambiguously
identify
the rank~3 winding sum 
with a minimal parabolic Eisenstein series at $\lambda_{32}=\lambda_{21}=1$.

\subsection{Integration over membrane volume -- non-degenerate orbit}
We now return to the issue of integrating over the membrane volume 
$\nu \in \Real^+$. As we argued above, the correct prescription is
to compute the constant term corresponding to the maximal parabolic group
$P_2$ generated by $(n_1,n_2)$. Physically, this amounts to computing
the Fourier coefficients associated to $(n_1,n_2)$ at zero momentum,
or equivalently, averaging over the action of $E_{\gamma_0}$ and
$E_{\om}$. Since $(y,x_0)$ are the conjugate variables 
(from~\eqref{sl3x1}), one should evaluate the spherical vector at 
the origin $y=x_0=0$. Unfortunately, this is the limit 
where the saddle point
approximation used to derive~\eqref{E6A2sph} breaks down. 
This is also the
limit in which the degenerate contributions are  obtained,
on which we have no control. 

A better strategy is therefore to intertwine back to the standard
induced representation~\eqref{sl3_cts}, for which constant terms
are completely known.  Having just established that the
$SL(3)_{NL}$ representation~\eqref{sl3x1} is equivalent to the 
induced representation $(-1,-1)$, it remains to establish how
the double pole at $(-1,-1)$ should be regularized. Unfortunately,
we have not been able to settle this point, however it is easy to find
an ad hoc prescription which gives the correct result: expanding
the constant term computation~\eqref{p2} around 
$\lambda_{21}=-1+\eps_1,\lambda_{32}=-1+\eps_2$, we have 
$$
E_{P_2} =
2t_1^{3-\eps_1-\frac12\eps_2} \zeta(2-\eps_2)
E^{SL(2)}_{{\bf 2},1-\frac12\eps_2}
+2t_1^{\frac32+\frac12\eps_1-\frac12\eps_2}\zeta(3-\eps_1-\eps_2)
\frac{\xi(1-\eps_1)}{\xi(\eps_1-1)}E^{SL(2)}_{{\bf
2},3/2-\frac12\eps_2-\frac12\eps_2} 
$$
\be
\ + \ 2t_1^{\frac12\eps_1+\eps_2}\zeta(2-\eps_1)
\frac{\xi(1-\eps_2)\xi(2-\eps_1-\eps_2)}
{\xi(\eps_2-1)\xi(\eps_1+\eps_2-2)}E^{SL(2)}_{{\bf
2},3/2-\frac12\eps_2-\frac12\eps_2}\, .
\ee
In this formula, the $\Real^+_\nu$ modulus  is related to
the membrane volume by $t_1=\nu^{1/6}$.
Picking the residue of the $1/\eps_2$ pole, extracting 
from it the finite term following the $1/\eps_1$ pole, and using
the relation $\zeta'(2)=-\zeta(3)/(4\pi^2)$, we find
\be
\lim_{\eps_1,\eps_2\to 0} \eps_2 \p_{\eps_1} \eps_1 E_{P_2} =
\frac{2\pi^3}{3\zeta(3}
{\widehat E}^{SL(2)}_{{\bf 2},1-\frac12\eps_1}
+\frac{2\pi^4}{3 \zeta(3)} \log t_1 + \frac{\pi^2}{3} t_1^3
+ \mbox{cste}
\ee
The volume independent term does indeed 
reproduce the $SL(2)$ Eisenstein series
from \eqref{fr4}! The meaning of the other terms is however far from
clear. It would be interesting to obtain a detailed understanding of
the singularity at $(\lambda_{32},\lambda_{21})=(1,1)$ from the
point of view of the membrane computation.
\section{Membranes on higher dimensional torii, membrane/5-brane duality}
\label{highertorii}

Having extolled our present understanding of the consequences
of the $E_6$ theta series conjecture for M-theory on $T^3$, 
we now briefly discuss the the higher dimensional generalization 
of our construction, and present a 
provocative hint that our framework may incorporate membrane/fivebrane 
duality.

\subsection{BPS membranes on $T^4$}
Let us first discuss the generalization of our construction to the
case of M-theory compactified on $T^4$--similar considerations 
can be applied to $E_8$ and $T_5$ compactifications. 
Here, we expect a symmetry
under the U-duality group $SL(5)$, which contains the obvious 
geometrical symmetry $SL(4)$ of the target 4-torus, together with
U-duality reflections which invert the volume of a sub 3-torus,
\be
R_{M} \to \frac{l_M^3}{R_N R_P}\ ,\quad
l_M^3 \to \frac{l_M^6}{R_M R_N R_P}
\ee
for any choice of 3 directions $(M,N,P)$ out of 4. The $R^4$ couplings
have been argued in~\cite{Kiritsis:1997em} to be given by an
$SL(5)$ Eisenstein series of weight 3/2. 

Just as for $T^3$, we expect this amplitude to be derivable
from a one-loop amplitude in membrane theory, {\it i.e.}, an integral
over the partition function  describing
minimal maps $T^3\to T^4$. In order to that the symmetry under
$SL(3)\times SL(5)$ be non-linearly realized, we assume an overarching
symmetry under $E_{7(7)}(\Zint)$ in the minimal representation,
which now has dimension 17. The canonical presentation 
of the minimal representation
 has a linearly realized $SL(6)$ subgroup,
under which the 17 variables are arranged as a $6\times6$ 
antisymmetric matrix $X$, and two singlets  $(y,x_0,X)$.
The spherical vector has been worked out 
in~\cite{Kazhdan:2001nx}, and, for large quantum numbers,
becomes an exponential of minus the action
\be
S=\frac{\sqrt{\det(X+|z|{\mathbb I}_6)}}{|z|^2}
 -i\ \frac{x_0{\rm Pf(X)}}{y|z|^2}\, ,
\ee
where $z\equiv y+ix_0$. 

This action however does not have a direct intepretation in terms 
of $T_4$ winding modes, which would make a $3\times 4$ matrix 
of integers. It is however possible to make a
judicious choice of polarization for the
Heisenberg subalgebra where $SL(3)\times SL(4)$ is linearly
realized, as follows. Let us examine the $E_6$ and $E_7$
extended Dynkin diagrams
\be
\begin{picture}(360,70)
\thicklines
\multiput(30,0)(30,0){4}{\circle{8}}
\put(0,-14){\makebox(0,0){$-\omega$}}\put(0,-25){\makebox(0,0){$$}}
\put(30,-14){\makebox(0,0){$\beta_0$}}\put(30,-25){\makebox(0,0){$$}}
\put(55.5,-3){$\times$}
\put(60,-12){\makebox(0,0){}}\put(60,-25){\makebox(0,0){$$}}
\put(90,-12){\makebox(0,0){}}\put(90,-25){\makebox(0,0){$$}}
\put(120,-12){\makebox(0,0){}}\put(120,-25){\makebox(0,0){$$}}
\multiput(34,0)(30,0){3}{\line(1,0){22}}
\put(60,4){\line(0,1){22}}
\put(60,30){\circle{8}}
\put(60,60){\circle{8}}
\put(60,34){\line(0,1){22}}
\put(6,0){\dots}\put(-5,-3){$\times$}

\multiput(214,0)(30,0){5}{\line(1,0){22}}
\multiput(210,0)(30,0){6}{\circle{8}}
\put(270,4){\line(0,1){22}}
\put(270,30){\circle{8}}
\put(295.5,-3){$\times$}
\put(180,-14){\makebox(0,0){$-\omega$}}\put(0,-25){\makebox(0,0){$$}}
\put(210,-14){\makebox(0,0){$\beta_0$}}\put(30,-25){\makebox(0,0){$$}}
\put(186,0){\dots}\put(175,-3){$\times$}

\end{picture}
\ee
\vspace{.4cm}

\noindent
The longest root $-\omega$ is denoted by a $\times$. The node labeled
$\beta_0$ determines the Levi subalgebra acting linearly on the Heisenberg
subalgebra (positions and momenta).
It is obtained by deleting nodes
$-\omega$ and $\beta_0$ 
which yields $SL(6)$ and $SO(6,6)$ groups, respectively.
The choice of a set of momenta and coordinates, {\it i.e.}, a polarization,
breaks these groups to a subgroup. The node marked with a cross
determines a choice of polarization appropriate for membrane winding
sums. The remaining nodes are subgroups linearly acting on positions, namely 
$SL(3)\times SL(3)$ and $SL(3)\times SL(4)$--precisely the linear
actions from left and right on 3~and~4-torus winding matrices.
Note that only for $E_6$ does the canonical polarization
determined by the node attached to $\beta_0$ coincide with the
membrane inspired one. For $E_8$, 5-torus winding sums, more general
choices of $\beta_0$ are even necessary.

Deleting the node marked with a cross from the extended Dynkin
diagrams, leaves the $U$-duality and world-volume groups.
We have discussed the $E_6$ case in detail. For $E_7$, 
the two rightmost nodes correspond to the $SL(3)$ membrane
world-volume shape moduli. The nodes remaining on the left form
$SL(6)\supset SL(5)\times \Real_+$, a product of the $T_4$  
$U$-duality group and the world-volume. (For $T_4$, the
integration over world-volumes is also regulated by quantum fluxes.) 
Also, just as  $E_6$ has a
triality, duality of the $E_7$ Dynkin diagram implies that there
is no ambiguity in the choice of linearly acting, $SL(3)$,  
world-volume modular group.

By looking at the root lattice, we may now determine the 
transformation properties of the seventeen
variables of the minimal representation with respect to the linearly
acting $SL(3)\times SL(4)$ subgroup: the singlet $y$ remains
unaffected, while $x_0$ and the elements of the $6\times 6$ 
antisymmetric matrix $X$ reassemble themselves into a
$4\times 3$ matrix $Z$ and a 4-vector $X$,
\be
X=\left(\begin{array}{c}X_{12}\\ X_{31}\\ X_{23}\\
  x_0\end{array}\right)
\ ,\qquad
Z=\left(\begin{array}{ccc}X_{14}&X_{15}&X_{16}\\
                       X_{24}&X_{25}&X_{26}\\
                       X_{34}&X_{35}&X_{36}\\
                       y\d_{X_{56}}&y\d_{X_{64}}&y\d_{X_{45}}
       \end{array}\right)\, .
\ee 
The derivatives in the last row indicate that a Fourier
transform should be performed in these variables before the symmetry
becomes linearly realized. The spherical vector in this 
new polarization now reads
\be
S=\frac{\sqrt{\det(Z^\rt Z+{\mathbb I}_{3\times3}(y^2+X_0^\rt X_0))
+{\cal R}/y^2
}}
{y^2+X_0^\rt X_0}+i\ \frac{\sqrt{\det(ZZ^\rt +X_0X_0^\rt )}}
{y(y^2+X_0^\rt X_0)}
\ee
where ${\cal R}$ stands for
\bea
{\cal R}&=&
(y^2+X_0^\rt X_0)
[(y^2+X_0^\rt X_0+\tr ZZ^\rt )\, X_0^\rt  ZZ^\rt  X_0- X_0^\rt ZZ^\rt ZZ^\rt X_0]\nn\\
&+&X_0^\rt X_0\, \det Z^\rt Z-\det(ZZ^\rt +X_0X_0^\rt )\, .
\eea
This action generalizes its $E_6$ counterpart~\eqn{action}. Again it
is an Born--Infeld-like 
interpolation between Nambu-Goto and Polyakov formulations. Over and
above the winding quantum numbers $Z$, additional fluxes $(y,X_0)$
are necessary for manifest $U$-duality invariance. Their
interpretation as a target space scalar and three-form also carries
through to the 4-torus case: the vector $X$ may be interpreted as
a 3-form flux on the world-volume, carrying also a 3-form index
in target space. It would be most interesting if the
integration over the world-volume $SL(3)$ shape moduli could be
carried out for this theory also.

\subsection{Membrane/five-brane duality and pure spinors}
\label{pure}
Going back to the $T^3,E_6$ case, and in line with the idea of
membrane/five-brane duality, it is now interesting to ask if a 
change of polarization might bring the minimal $E_6$ theta series to
a form which could be interpreted as a five-brane partition 
function\footnote{We are grateful to L. Motl for forcing
us to address this question.}.

Indeed, as noticed in~\cite{Kazhdan:2001nx}, it is possible to 
choose a polarization where $SL(5)$ becomes linearly realized, instead
of $SL(3)\times SL(3)$. By Fourier transforming the one-row of the 
$3\times 3$ matrix, the 11 variables $(y,x_0,Z^M_A)$ rearrange
themselves into a $5 \times 5$ antisymmetric matrix $X$,
\be
X = \begin{pmatrix} 0 & -\p_{Z_3^3} & \p_{Z_3^2} & Z_1^1 & Z_2^1 \\
 & 0 & - \p_{Z_3^1} & Z_1^2 & Z_2^2 \\
 &   & 0 &  Z_1^3 & Z_2^3 \\
& a/s & & 0 & x_0 \\
&&&&0
\end{pmatrix}
\ee
and a singlet $y$. In fact, these 11 variables can be supplemented by
a 5-vector $v_i$, expressed in terms of the others as 
$y v^i=\frac14 \eps^{ijklm} X_{ij} X_{kl}$, in such a way that the 16
variables $(y, X_{ij}, \eps_{ijklm} v^m)$ transform as a
Majorana--Weyl
spinor $\lambda$
of $SO(5,5) \subset E_6$, subject to the ``pure spinor'' constraint 
$\lambda \Gamma^\mu \lambda=0$, for all $\mu = 1\dots 10$.
This is the direct analog for $E_6$ of the ``string inspired''
representation of $D_4$ on 6 variables $m^{ij}$, subject to the
quadratic constraint $\eps_{ijkl} m^{ij} m^{kl}=0$. A completely 
analogous construction for the $E_7$ minimal
representation holds also, see~\cite{leshouches} for details.

In this polarization, the $E_6$ spherical vector now takes the very
simple form\footnote{Minimal representations and their
spherical vectors have been studied before in this presentation
in~\cite{dvorsky}.}, 
\be
f_{E6}={\cal K}_1\Big(\sqrt{\overline\lambda\lambda}\Big)\, ,\qquad
{\cal K}_t(x)\equiv x^{-t}K_t(x)\, .
\ee
We now claim that this can be interpreted as the partition function
for the five-brane. Indeed, we can view the $5\times 5$ antisymmetric
matrix $X$ as the electric flux on the five-brane, $X_{ij}=H_{0ij}$,
equal by the self-duality condition to the magnetic components,
$X^{ij} = (1/6) \eps^{ijklm} H_{klm}$. An $SL(6)$-invariant
partition function for fluxes on a single five-brane
has been constructed long ago
by Dolan and Nappi in~\cite{Dolan:1998qk}, where the $SL(6)$
invariance was non-linearly realized thanks to Gaussian
Poisson resummation -- in other words, the $SL(6)$
symmetry was embedded in an overarching symplectic group $Sp(10)$
acting linearly on the 10 variables $X_{ij}$. Here, we find that
by adding an extra variable $y$, best thought of as the number of 
five-branes, the geometric symmetry $SL(6)$ is enlarged to
a complete $E_{6(6)}$ group. This is the symmetry group 
expected for a five-brane wrapped on a $T^6$ with metric
$g_{ij}$, 3-form $C_{ijk}$ and dual 6-form $E_{ijklmn}$ parameterized
by the symmetric space $E_{6(6)}/Usp(4)$. 
It would be very interesting to use this information to compute
the partition function for $y$ overlapping five-branes.

Independently of these mem- and five-brane considerations, 
let us note that pure spinors of $D_5$ (albeit with a different
real form, $SO(1,9)$) arise naturally in the description 
of 10-dimensional Yang-Mills theory: 
as observed by Berkovits~\cite{Berkovits:2000fe}, 
the $N=1$ superalgebra in ten dimensions
\be
\{Q_\alpha,Q_\beta\}=\Gamma^\mu_{\alpha\beta} P_\mu\, ,
\ee
becomes cohomological when  auxiliary, pure spinor variables
$\lambda^\alpha$ obeying
\be
\lambda \Gamma^\mu \lambda=0\, ,\label{10=5}
\ee
are introduced, because one can then form the nilpotent operator
\be
{\cal Q}\equiv\lambda^\alpha Q_\alpha\, .
\ee
Superfields are then also functions of pure spinors $\lambda^\alpha$. 
Here, we have shown that the space of functions of $\lambda^\alpha$ is
precisely the minimal representation of $E_6$. A tantalizing
possibility,
therefore, is a hidden $E_6$ symmetry of $d=10$ super Yang-Mills theory.

\section{Discussion}
\label{discussion}

In this Article, we have attempted to compute the $R^4$ coupling
in M-theory compactified on $T^3$, 
from a one-loop computation in the microscopic 
membrane theory. Our basic assumption has been that, at least
at the zero-mode level, the membrane theory enjoys an ``overarching'' 
symmetry under the minimal representation $E_{6(6)}(\Zint)$, 
which mixes world-volume and target space. This provides the most
economical way to realize both the modular symmetry $SL(3,\Zint)$ on
the world-volume, and U-duality $SL(3,\Zint) \times SL(2,\Zint)$
in target space. An important consequence is
that there should be two discrete degrees of freedom, $y$ and $x_0$,
over and above the nine evident winding numbers $Z_A^M=\p_A X^M$. 
After integrating over the worldvolume moduli, only the U-duality
symmetry remains. 

Using explicit computations and representation
theoretic arguments, we have found that the rank$(Z)=1$ and rank$(Z)=2$
contributions were equal, and reproduced the $SL(3)$ part of the
$R^4$ coupling in~\eqref{fr4}, while the rank 0 and 3 contributions
ought
reproduce the $SL(2)$ part, {\it i.e.}, the membrane instanton sum. 
A fully satisfying proof would require an understanding of the
regularization of the poles of the Eisenstein series at
$(\lambda_{21},\lambda_{32})=(1,1)$, which we have not been 
able to obtain thus far. A possible avenue to this end
would be to derive the exact expression for the integrated spherical 
vector beyond the steepest descent approximation~\eqref{E6A2sph},
and study the limit $(y,x_0)\to (0,0)$. This would shed much
desired light on the interpretation of the new quantum numbers
$(y,x_0)$, which has remained rather  elusive.

On this subject, we may offer the following observations:
the $E_6$ and $E_7$ cases suggest
that the singlet $x_0$ (resp. the 4-vector $X$) should correspond
to a flux carrying 3-form index both on the membrane world volume.
On the other hand,  rescaling the
winding variables $Z\rightarrow yZ$ and also $x_0\rightarrow yx_0$,
we observe that the variable $y$ appears linearly as an overall factor in the
action~\eqn{action}. Therefore, it is natural to identify it 
with the number of membranes. It would be interesting to understand
how to introduce these degrees of freedom in an off-shell
membrane action, which would realize the overarching symmetry
non-linearly. Equation~\eqn{action} does at least predict the shape of 
the action of such a theory, restricted to zero-modes. The putative
appearance of a membrane number, even calls for a second
quantized membrane field theory.

Another interesting question is how does the double dimensional 
reduction of a membrane winding sum to the $T$-duality invariant
string amplitude work? At the level of the final amplitude~\eqn{fr4},
this is no mystery since this is essentially
how it was conjectured in the first
place~\cite{Kiritsis:1997em}. However, the string amplitude is based
on a $T$-duality/world-volume modular group,
dual pair construction for the symplectic groups. An appealing, but
{\it not} necessary, scheme would be to embed these in the exceptional
groups proposed here. Unfortunately this seems not to work, since
a quick perusal of the branching rules for the $E_6$ does not yield
the requisite $Sp(8)$ subgroup.

Finally, we have presented two observations that reach beyond
the subject of supermembranes: first, that the minimal representation
of $E_6$ can also be used to describe stacks of $N$ 
five-branes wrapped on $T^6$, in a manifestly U-duality invariant
way: it would be very interesting to use this result to compute
five-brane partition functions at finite $N$. Second, pure
spinors in ten dimensions carry an action of $E_6$: the implications
for 10-dimensional Yang-Mills theory remain to be uncovered.

\section*{Acknowledgments}

It is a pleasure to thank David Kazhdan and Stephen Miller, 
our mathematical muses, for
extensive consultations. We would also like to acknowledge 
Colette Moeglin, Lubos Motl, Misha Movshev, Hermann Nicolai, 
Jan Plefka and Alexander Polishchuk for useful discussions.
We thank the Albert Einstein Institut f\"ur Gravitationsphysik and
the Jefferson Physical Laboratory at Harvard University
for kind hospitality. A.W. thanks also Max Planck Institut f\"ur
Mathematik and LPTHE Universit\'e Paris~6 et~7 for
hospitality. Research 
supported in part by NSF grant PHY01-40365
and the European Network HPRN-CT-2000-00131.

\begin{appendix}

\section{Summation measure and degenerate contributions}

\label{measdeg}

It is well known that the summation measure of a theta series
may be expressed adelically, {\it i.e.}, as a product over primes
of the spherical vector over the $p$-adic base field 
(see {\it e.g.}~\cite{leshouches} for a physicist's review)
\be
\mu(y,x_0,Z)=\prod_{p\,  \rm prime} f_{\Q_p}(y,x_0,Z)\, .
\ee
The function $f_p$ is determined by requiring invariance under
the maximal compact subgroup $Usp(4,\Q_p)$.
Its explicit form can be obtained by noting that the exponent
or ``action'' $S$
in~\eqn{theta} has a simple rewriting in terms of the Euclidean norm
on a certain Lagrangian submanifold. Some details: Consider the space
of coordinates $X=(x_0,y,Z)$ along with canonical momenta $P$. Introduce
a ``Hamiltonian'' 
\be
H=\frac{\det Z}{||(y,x_0)||}\, .
\ee
For now $||v||\equiv \sqrt{v.v}$ denotes the Euclidean norm of a
vector $v$. The Lagrangian submanifold is described by the subspace
$(X,P=\nabla_X H)$. The action then has a very simple rewriting
\be
S=||(X,\nabla_X H)||+i\frac{x_0\, H}{y \, ||(y,x_0)||}\ . 
\ee
Recalling that the exponential form of the $E_6$
spherical vector arose because of the identity
$K_{1/2}(x)=\sqrt{\frac{\pi}{2x}} \exp(-x)$ we may write
\be
\label{fsr}
f^{E_6}_{\Real}=
\frac{
{\cal K}_{1/2}({\rm Re}(S))e^{-i{\rm Im}(S)}
}{||(y,x_0)||^2}\, ,
\ee
where ${\cal K}_t(x)\equiv x^{-t}K_t(x)$.

The $p$-adic spherical vector was obtained recently in~\cite{KazPol},
and takes a very similar form. In the $E_6$ case, it reads,
for $||y||_p \leq ||x_0||_p$,
\be
f_{\Q_p}^{E_6} =||x_0||_p^{-2}\,
\frac{p\, ||(x_0,Z,\nabla_Z \frac{\det Z}{x_0},\nabla_{x_0} \frac{\det Z}{x_0} )||_p^{-1}-1}{p-1}\,  \exp\left( - i \frac{\det Z}{y x_0} \right)
\ee
if $||(x_0,Z,\nabla_Z \frac{\det Z}{x_0},\nabla_{x_0} \frac{\det Z}{x_0} )||_p\leq 1$, and 0 otherwise.
The opposite case $||y||_p > ||x_0||_p$ is simply obtained by a 
Weyl reflection, the effect of which is  to  eliminate the phase and
exchange $y$ and $x_0$. Notice that the $p$-adic spherical
vector can be simply obtained from the real one 
by replacing the orthogonal
norm $||.||^2$ by the $p$-adic norm and the function  ${\cal
K}_{1/2}$ by its simple (algebraic) $p$-adic analog.
The summation measure is the product of $p$-adic spherical vectors 
over all prime $p$, therefore has support on rational numbers $X$
such that, for all prime $p$, either $||y||_p \leq ||x_0||_p$ and $||(x_0,Z,\nabla_Z \frac{\det Z}{x_0},\nabla_{x_0} \frac{\det Z}{x_0} )||_p\leq 1$, or $||y||_p > ||x_0||_p$ and $||(y,Z,\nabla_Z \frac{\det Z}{y},\nabla_{y} \frac{\det Z}{y} )||_p\leq 1$. Thus, the summation measure has support 
on integers $(y,x_0,Z)$ such that 
$\gcd(y,x_0)$ divides $\nabla_Z (\det Z)$ and $(\gcd(y,x_0))^2$ divides $\det Z$.

The degenerate terms
for exceptional theta series have also been determined in~\cite{KazPol}. 
These terms are familiar to
physicists in the context of the large volume expansion of the well 
known non-holomorphic $SL(2)$ Eisenstein series, see for 
example equation~\eqn{instantons}.
The explicit
expression for the $E_6$ minimal theta series given in~\eqn{theta} is the analog of
the bulk membrane instanton term in~\eqn{instantons}.
In~\cite{KazPol}, the
degenerate terms for the $E_6$ theta series are found to be
\be
\theta_{E_6}=\!\sum_{\stackrel{y\in\Zint\setminus\{0\}}{\sss (x_0,Z)\in
\Zint^{10}}}\left[\prod_{p\ \rm prime}\ f^{E_6}_{\Q_p}(y,x_0,Z)\right] 
f^{E_6}_{\Real}(y,x_0,Z)
+\sum_{(x_0,Z)}\mu(x_0,Z)\overline{f}(x_0,Z)+\alpha_1+\alpha_2\, .
\label{KPformula}
\ee
The leading term is the one above in~\eqn{theta}. As exploited in a 
cosmological
context in~\cite{Pioline:2002qz}, the minimal representation has a natural
conformal symmetry, the underlying geometry of the
eleven dimensional space with variables $(y,x_0,X)$ being conical.
The second term in~\eqn{KPformula} corresponds to the limit
$y\rightarrow0$
at the tip of the cone: the regulated spherical vector
$\overline f$ is obtained by dropping the phase factor in~\eqref{fsr},
and setting $y=0$ in the remaining part.
 The final two terms $\alpha_1$ and $\alpha_2$
are automorphic forms
built from singlet and minimal representations of the Levi subgroup
$SL(6)\subset E_6$. Again, for details we refer the reader 
to~\cite{KazPol}.

\section{$SL(3,\Zint)$ automorphic forms}

\label{AutoApp}

A most useful account of $SL(3,\Zint)$ automorphic forms may be
found in~\cite{Miller}. In particular, 
the constant term computations presented in Section~\ref{constant_term_app}
are derived in detail there. It was realized long ago, especially by
Langlands, that representation theory plays a central r\^ole in the 
theory of automorphic forms. Therefore in Section~\ref{sl3-rep}
we construct the continuous series irreducible representations of $SL(3)$ upon
which $SL(3,\Zint)$ invariant automorphic forms are based.
We review the fundamental domain construction for $SL(3,\Zint)$ in 
Section~\ref{FDSL3} (a computation first tackled many years ago by
Minkowski!). The remaining sections deal with $SL(3)$ Eisenstein 
series and their constant terms.

\subsection{Continuous representations by unitary induction}
\label{sl3-rep}
All continuous unitary irreducible representations of $SL(3)$ have been 
constructed long ago in~\cite{Vahut}, by unitary induction 
techniques. According to the general
philosophy of Kirillov~\cite{Kirillov}, 
they are in one-to-one correspondence with coadjoint
orbits. We recall these results here for convenience.

$SL(3)$ admits a parabolic subgroup $P$ of upper triangular matrices
\be
\label{indp}
P\equiv\left\{p=
\left(
\begin{array}{ccc}
t_3&*&*\\
0&t_2&*\\
0&0&t_1
\end{array}
\right) \, :\ t_{1}t_{2}t_{3}=1\right\}\subset SL(3)\, ,
\ee
A one-dimensional representation of $P$ is given by the character
\be
\label{char}
\chi(p,\lambda) = \prod_{i=1}^3 {\rm sgn}^{\epsilon_i}(t_{i})\, 
|t_{i}|^{\rho_i}\, .
\ee 
Here $\epsilon_i=\pm 1$ and the three complex variables $\rho_i$
are conveniently parameterized as
\be
\rho_1=\lambda_1+1\, ,\quad \rho_2=\lambda_2\, ,\quad
\rho_3=\lambda_3-1\, ,
\ee 
where the constants represent a shift by the longest root.
The unit determinant property implies that all results are written in 
terms of differences
\be
\lambda_{ij}\equiv \lambda_i-\lambda_j\, .
\ee
We will often use the notation $\lambda=(\lambda_{32},\lambda_{21})$.
Functions on
$SL(3)$ which transform by a factor $\chi(p,\lambda)$ 
under left multiplication 
by $p_0 \in P$ yield a continuous irreducible representation of 
$SL(3)$. Functions with the above covariance reduce to functions
of the three coordinates  $(x,v,w)$ of the coset $P\backslash SL(3)$ 
parameterized by the gauge choice 
\be
g = 
\begin{pmatrix}
1    &   & \\
x    & 1 & \\
v+xw & w &\, 1\,
\end{pmatrix}\, ,
\label{para_rep}
\ee
suitably extended to $SL(3)$ via the character $\chi$. The
action of the infinitesimal generators of $SL(3)$ on functions
$f(x,v,w)$ is computed straightforwardly, and reads
\bea
\begin{array}{lr}
E_{\g}=-\partial_x +w \partial_v \ , &
E_{-\g}=-x^2\partial_x-v\partial_w+(\lambda_{32}-1)x \, ,\\[1mm]
E_{\b}=\partial_w \ , \qquad&\hspace{-1cm}
E_{-\b}=w^2\partial_w+vw\partial_v-(v+xw)\partial_x
+(1-\lambda_{21})w\, ,\\[1mm]
E_{\omega}=\partial_v \ , &\hspace{-5.0cm}
E_{-\omega}=v^2\partial_v+vw\partial_w+x(v+xw)\partial_x 
-(\lambda_{31}-2)v-\lambda_{32} xw\, ,\\[1mm]
H_\gamma=2x\partial_x+v\partial_v-w\partial_w-(\lambda_{32}-1)\, , &
H_\beta=-x\partial_x+v\partial_v+2w\partial_w-(\lambda_{21}-1)\, .
\end{array}\nn\\
\label{sl3_cts}
\eea
(Note that the discrete choices of $\epsilon_i$ are not visible
at the infinitesimal level.) 
Let us now compute the quadratic and cubic Casimir invariants for this
representation as given in~\eqn{casimirs}:
\bea
{\cal C}_2&=&\frac16\ \big( 
\lambda_{21}^2+\lambda_{32}^2+\lambda_{13}^2 \big) -1\, , \\
{\cal C}_3&=&\frac12\ (\lambda_{13}-\lambda_{32})(\lambda_{21}-\lambda_{13})
(\lambda_{32}-\lambda_{21})\, .
\eea
We will be particularly interested in the representations such
that ${\cal C}_2={\cal C}_3=0$.
These correspond to $(\lambda_{32},\lambda_{21})=
(1,1),(2,-1),(1,-2),(-1,-1),(-2,1),
(-1,2)$ and appear at the six intersection of the radial lines with
the ellipse depicted in Figure~\ref{sl3cas}. 
\FIGURE{ \label{sl3cas}
\hfill\epsfig{file=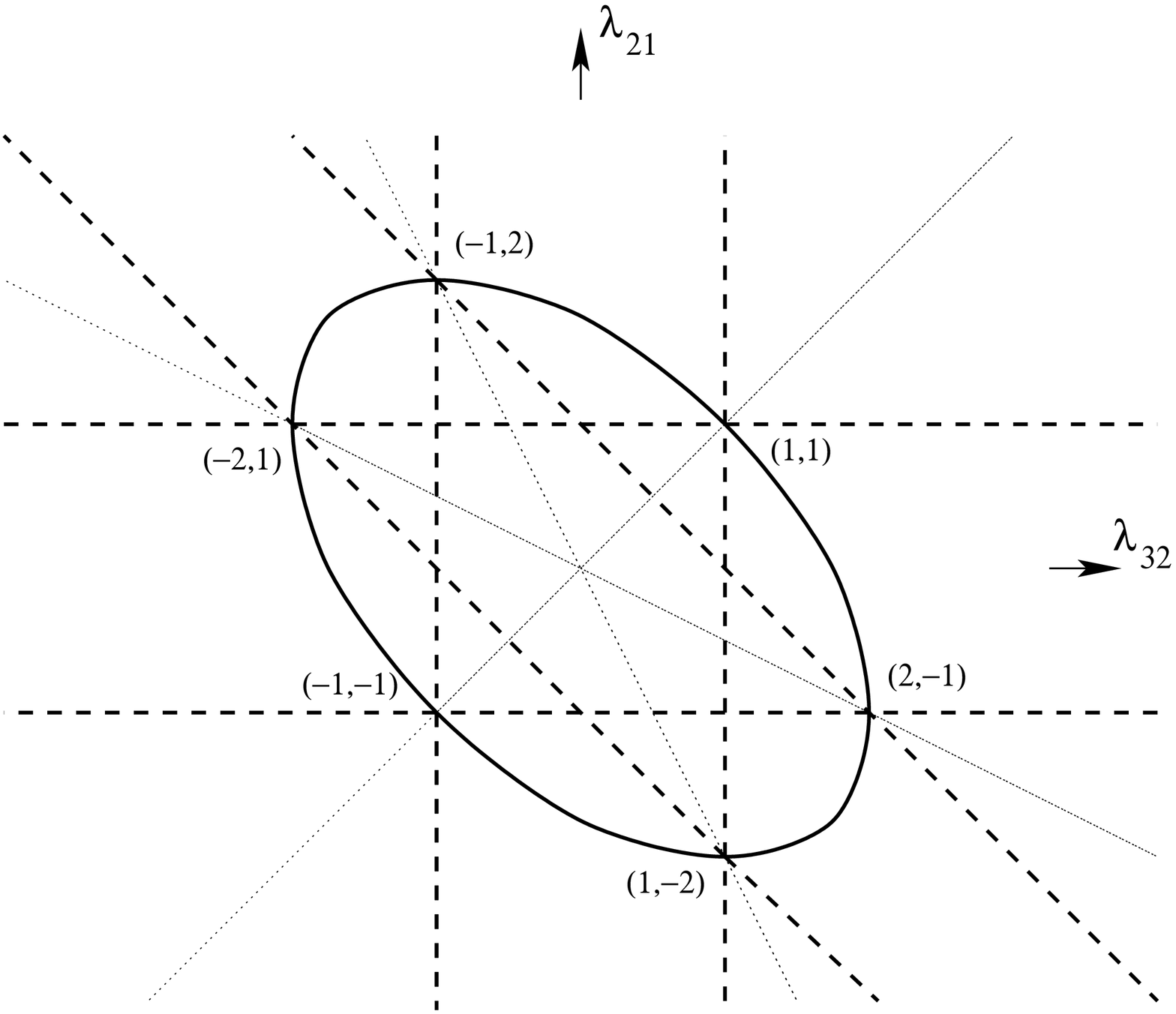,height=8cm}\hfill
\caption{Vanishing loci of ${\cal C}_2, {\cal C}_3$ in the 
$(\lambda_{32},\lambda_{21})$ plane. The quadratic Casimir vanishes
on the ellipse, the cubic on the dotted radial lines, while the dashed 
lines correspond to simple poles of Eisenstein series. 
}}

The spherical vector in this representation is readily found:
The maximal compact subgroup $SO(3)\subset SL(3)$ acts on the columns
of the coset representative $g$ in~\eqn{para_rep} as rotations of 3-vectors
while the compensating $P$ gauge transformation rescales the first column
$(1,x,v+x w)$ by $t_{3}$ and the wedge product of first and second rows
$(-v,w,1)$ by $t_{3}t_{2}$. Hence the product of norms raised
to the appropriate powers
\be
f_{SL(3)}=\left[1+x^2+(v+xw)^2\right]^{-s} 
\ \left[1+v^2+w^2\right]^{-t}\, ,\nn\\
\label{spherical_A2}
\ee
with
\be
s\equiv\frac{1}{2}\ \big(1-\lambda_{32}\big)\ ,\quad
t\equiv\frac{1}{2}\ \big(1-\lambda_{21}\big)\, ,
\ee
is simultaneously invariant under the right action of $P$ 
and left action of $SO(3)$ and is therefore the desired spherical vector.

It is also interesting to obtain the minimal representation from the
limit in which the induced representation~\eqn{sl3_cts} approaches any
one of the simple pole dashed lines in Figure~\ref{sl3cas}. For example,
setting $\lambda_{32}=1$ and $\lambda_{21}=1-2t$, the induced
representation acts faithfully on functions $\phi(v,w)$, by setting
$\partial_x=0$. We then find
\bea
\begin{array}{lr}
E_{\g}=w \partial_v \ , &
E_{-\g}=-v\partial_w \, ,\\[1mm]
E_{\b}=\partial_w \ , \qquad&
E_{-\b}=w^2\partial_w+vw\partial_v
+2tw\, ,\\[1mm]
E_{\omega}=\partial_v \ , &
E_{-\omega}=v^2\partial_v+vw\partial_w
-2tv\, ,\\[1mm]
H_\gamma=v\partial_v-w\partial_w\, , \qquad\qquad\qquad&
H_\beta=v\partial_v+2w\partial_w+2t\, .
\end{array}\nn\\
\label{sl3_min1}
\eea
The spherical vector of this representation is simply
$f_{SL(3),{\rm min}}=[1+v^2+w^2]^{-t}$ corresponding
to an Eisenstein series $E_{{\bf 3},t}^{SL(3)}$
in the fundamental representation. 
Another useful presentation of
the minimal representation, used in the Text, is obtained by
Fourier transform over $(v,w)\rightarrow\frac1i(\d_v,\d_w)$.
In this case we find
\bea
\begin{array}{lr}
E_{\g}=-v \partial_w \ , &
E_{-\g}=w\partial_v \, ,\\[1mm]
E_{\b}=-iw \ , \qquad&
E_{-\b}=iw\partial_w^2+3i\partial_w+iv\partial_w\partial_v
-2it\partial_w\, ,\\[1mm]
E_{\omega}=-iv \ , &
E_{-\omega}=iv\partial_v^2+3i\partial_v+iw\partial_v
\partial_w
-2it\partial_v\, ,\\[1mm]
H_\gamma=-v\partial_v+w\partial_w\, , \qquad\qquad\qquad&
H_\beta=-v\partial_v-2w\partial_w+2t-3\, .
\end{array}\nn\\
\label{sl3_min2}
\eea
The spherical vector is then
\be
\widetilde f_{SL(3),{\rm min}}
={\cal K}_{1-t}\Big(\sqrt{v^2+w^2}\Big)\, ,\qquad\quad
{\cal K}_t(x)\equiv x^{-t}K_t(x)\,.
\label{fourminsph}
\ee
This form corresponds to the non-degenerate terms in a large
volume expansion of the Eisenstein series in the
fundamental representation~\cite{Kazhdan:2001nx}.

\subsection{Fundamental domain of $SL(3,\Zint)$}
\label{FDSL3}

The fundamental domain of $SL(3,\Zint)$ in $SO(3)\backslash SL(3,\Real)$
was first computed by Minkowski~\cite{Minkowski}, here we follow
the modern treatment of~\cite{Grenier}.
We parameterize the coset $SO(3)\backslash SL(3,\Real)$ 
in the Iwasawa gauge as,
\bea
e&=&
\begin{pmatrix} \frac{\ts 1}{\ts  L} && \\ 
&\sqrt{\frac{\ts  L}{\ts T_2}} & \\ 
&& \sqrt{ LT_2}
\end{pmatrix} \cdot
\begin{pmatrix} \ 1\  & A_1 & A_2 \\[2mm]
& 1 & T_1 \\[2mm] 
& & \ 1\ \ 
\end{pmatrix}
\equiv
\frac1 L  \begin{pmatrix} 
\vec e_1 & \vec e_2 & \vec e_3
\end{pmatrix}
\eea
where we defined the three vectors
\bea
\vec{e}_1=\begin{pmatrix} 1 \\ 0 \\ 0 \end{pmatrix} \, , \quad
\vec{e}_2=
\begin{pmatrix} A_1 \\ \sqrt{\frac{\ts  L^3}{\ts T_2}} \\ 0 
\end{pmatrix}\,  ,
\quad
\vec{e}_3=
\begin{pmatrix} A_2 \\ \sqrt{\frac{\ts  L^3}{\ts T_2}} \ T_1\\ 
\sqrt{ L^3T_2} 
\end{pmatrix}\,  .\qquad
\eea
The fundamental domain is constructed by 
maximizing a height function
\be
h(e)\equiv (\vec e_1)_1 \ (\vec e_2)_2\  (\vec e_3)_3= L^3\, .
\ee
The function $h$ is invariant under the right action of
the following $SL(3,\Zint)$ elements\footnote{Note that the set of
elements $\{S_{1,2,3,4,5},T_{1,2,3},U_{1,2}\}$ is convenient for
determining the fundamental domain, yet overcomplete--$PSL(2,\Zint)$
is generated by the shift and clock matrices\vspace{-.18cm}
$$\begin{pmatrix}1&&1\\&\, 1\, &\\&&1\end{pmatrix}\;\; \mbox{ and }\;\;
\begin{pmatrix}
&\, 1\, &\\&&1\\1&&\end{pmatrix}\, .$$\vspace{-.19cm}}
\be
{\rm S}_2=\begin{pmatrix}  1&& \\ && \,\, 1\\&-1&  \end{pmatrix}\, ,\quad
{\rm T}_1=\begin{pmatrix}  1&\ 1\ & \\ &1&\\&&1  \end{pmatrix}\, ,\quad
{\rm T}_2=\begin{pmatrix}  1&&1 \\ &\ 1\ &\\&&1  \end{pmatrix}\, ,\quad
{\rm T}_3=\begin{pmatrix}  1&& \\ &\ 1\ &1\\&&1  \end{pmatrix}\, ,\quad
\nn
\ee
\be
{\rm U}_1=\begin{pmatrix}  -1&& \\ &-1&\\&&\ 1\   \end{pmatrix}\, ,\quad
{\rm U}_2=\begin{pmatrix}  1&& \\ &-1&\\&&-1  \end{pmatrix}\, ,\quad
\label{STU1}
\ee
Therefore, we can always move to a fundamental domain
of the $Gl(2,\Zint)$ in the bottom right hand
$2\times2$ block along with strip conditions on the $A_{1,2}$,
\be
T_1^2+T_2^2\geq1\, , \qquad
0\leq A_1,T_1\leq \frac12\, ,\qquad
-\frac12\leq A_2\leq \frac12\, .
\label{FD1}
\ee
There are however additional conditions\footnote{Some of these were missed
in~\cite{Pioline:2001jn}, although this omission does not alter
the conclusions found there.} following from the
actions of further $SL(3,\Zint)$ elements on the height function
\bea
{\rm S}_1=\begin{pmatrix}  &  1 & \\ -1 & & \\ & &\,\, 1\   \end{pmatrix}
&:&h\mapsto \frac{h}{|\vec e_2|^3}\nn\\
{\rm S}_3=\begin{pmatrix} && 1\ \\ & 1& \\-1&&  \end{pmatrix}
&:&h\mapsto \frac{h}{|\vec e_3|^3}\nn\\
{\rm S}_4=\begin{pmatrix} &\ 1\ & \\ \ \ 1& &-1 \\-1&&  \end{pmatrix}
&:&h\mapsto \frac{h}{|\vec e_3-\vec e_2|^3}\nn\\
{\rm S}_5=\begin{pmatrix}\ \ 1 && \\ -1&\ 1\  & \\\ \ 1&&1\   \end{pmatrix}
&:&h\mapsto \frac{h}{|\vec e_3-\vec e_2+\vec e_1|^3}\, .
\label{STU2}
\eea
Clearly\footnote{The proof: Without loss of
generality assume $A_1\geq 0$. Then if $|\vec e_2|<1$, act with ${\rm S}_1$
to increase~$h$. Since ${\rm S}_1:|\vec e_2|\mapsto 1/|\vec e_2|$, 
further actions
of ${\rm S}_1$ do not increase~$h$.
Next act with ${\rm S}_3$ in the case that $|\vec e_3|<1$. A single action
suffices because ${\rm S}_3:|\vec e_3|\mapsto 1/|\vec e_3|$.
Furthermore, since ${\rm S}_3:|\vec e_2|\mapsto |\vec e_2|/|\vec e_3|$
this does not disturb the 
previous $|\vec e_2| \geq 1$ inequality. Similarly, if $|\vec e_3-\vec
T|<1$, 
act with ${\rm S}_4$, which does not ruin previous equalities as
${\rm S}_4:(|\vec e_2|,|\vec e_3|)\mapsto(1/|\vec e_3-\vec e_2|,|\vec
e_2|/|\vec e_3-\vec e_2|)$. 
Also, only one ${\rm S}_4$ action is needed because ${\rm S}_4:|\vec e_3-\vec e_2|
\mapsto |\vec e_1+\vec e_2|/|\vec e_3-\vec e_2|$ and
$|\vec e_1+\vec e_2|\geq1$ for $A_1\geq 0$.
Finally, when $|\vec e_3-\vec e_2+\vec e_1|<1$, act with ${\rm S}_5$.
This respects previous inequalities because
${\rm S}_5:(|\vec e_2|,|\vec e_3|,|\vec e_3-\vec e_2|)\mapsto
(|\vec e_2|/|\vec e_3-\vec e_2+\vec e_1|,
|\vec e_3|/|\vec e_3-\vec e_2+\vec e_1|,
|\vec e_3-\vec e_2|/|\vec e_3-\vec e_2+\vec e_1|)$.
Only finitely many applications of ${\rm S}_5$ are necessary because
${\rm S}_5:|n(\vec e_3-\vec e_2) +\vec e_1|\mapsto
|(n+1)(\vec e_3-\vec e_2) +\vec e_1|/|\vec e_3-\vec e_2+\vec
e_1|$. {\bf QED} 
} 
 maximal $h$ requires 
$|\vec e_2|,|\vec e_3|,|\vec e_3-\vec e_2|,|\vec e_3-\vec e_2+\vec
e_1|\geq 1$, {\it i.e.}, 
\bea
1&\leq& A_1^2+\frac{ L^3}{T_2}\nn\\
1&\leq& A_2^2+\frac{ L^3}{T_2}\ T_1^2 + L^3T_2\nn
\eea
\bea
1&\leq& (A_2-A_1)^2+\frac{ L^3}{T_2}\ (T_1-1)^2 
+ L^3T_2\nn\\
1&\leq& (A_2-A_1+1)^2+
\frac{ L^3}{T_2}\ (T_1-1)^2 + L^3T_2\, .
\label{FD2}
\eea
Conditions~\eqn{FD1} and~\eqn{FD2} constitute the fundamental domain
since the (non-minimal) set of elements considered generate $SL(3,\Zint)$.
Finally, note that in these variables,
the $SL(3)_L$ invariant integration measure is 
\be
de=\frac{d^2T}{T_2^2}\ d^2\!A\ \frac{d L}{ L^4}\, .
\ee 

\subsection{$SL(3)$ Eisenstein Series}
\label{eis_app}

In this Appendix we review prescient aspects of the
theory of Eisenstein series and gather formulae
needed in the main text for $SL(3)$. For our
purposes it suffices to consider Eisenstein
series based on the minimal parabolic subgroup
\be
P=\left\{\begin{pmatrix}*&*&*\\{}&*&*\\&&*\end{pmatrix}\right\}
\subset SL(3,\Real)\, ,
\ee
which we decompose into its nilpotent radical and Levi components
\be
P=MN\, ,\qquad
M=\left\{\begin{pmatrix}*&&\\&*&\\&&*\end{pmatrix}\right\}\, ,\qquad
N=\left\{\begin{pmatrix}1&*&*\\{}&1&*\\{}&&1\end{pmatrix}\right\}\, .
\ee
A general element $g\in SL(3,\Real)$ has the Iwasawa decomposition
\be
g=
k\ \begin{pmatrix}t_3&&\\&t_2&\\&&t_1\end{pmatrix}
\begin{pmatrix}1&a_3&a_2\\&1&a_1\\&&1\end{pmatrix}
\ee
where $t_1t_2t_3=1$ and $k\in SO(3)$ the maximal compact subgroup of
$SL(3,\Real)$. We can then introduce the function on $SL(3)$
\be
F(g;\lambda)\equiv t_1^{\lambda_1+1}t_2^{\lambda_2} t_3^{\lambda_3-1}
=t_2^{\lambda_{21}-1}t_3^{\lambda_{31}-2}\, ,
\qquad \lambda_{ij}\equiv\lambda_i-\lambda_j,
\ee
which is manifestly invariant under left action of the compact
$SO(3)$ and right action of the nilpotent $N$. It is also an
eigenfunction of the $SL(3,\Real)$-invariant differential operators 
$\Delta^{(2,3)}_{SL(3)}$ introduced in Section~\ref{cassies}.

To produce an automorphic form, 
we average (automorphize) $F(g;\lambda)$ over the right action of 
$SL(3,\Zint)$. Indeed since $F(g;\lambda)=F(gn;\lambda)$ for $n\in N$
it suffices to sum over $N(\Zint)\backslash SL(3,\Zint)$ 
(where $N(\Zint)\equiv SL(3,\Zint)\cap P$). Therefore, the $SL(3)$
Eisenstein series for the minimal parabolic is defined
as
\be
E(g;\lambda)=\sum_{\gamma\in N(\Zint)
\backslash SL(3,\Zint) } F(g\gamma;\lambda)\, .
\label{Eisenstein}
\ee 
This is now manifestly invariant under left action of $SO(3)$ and
right action of $SL(3,\Zint)$, therefore an automorphic function
on $SO(3)\backslash SL(3)$. As usual, this automorphic form can
be decomposed as in~\eqref{formula}, where the representation 
$\rho$ is precisely the representation induced from the minimal
parabolic $P$ with character~\eqref{char}\footnote{Indeed, the
right action on $g$ can be converted into a left action on
$\gamma$, which takes value in $P\backslash SL(3)$.
Eisenstein series for the minimal parabolics are defined
in a similar way. For example, for $P_1$ defined in~\eqn{parabolics}, 
elements of $SL(3,\Real)$ are decomposed as 
$$  g=k
\begin{pmatrix}t_3&&\\&\frac{1}{\sqrt{t_3}}&\\&&\frac{1}{\sqrt{t_3}}
\end{pmatrix}
\begin{pmatrix}1&&\\&\frac{1}{\sqrt{\tau_2}}&\\&&\sqrt{\tau_2}
\end{pmatrix}
\begin{pmatrix}1&&\\&1&\tau_1\\&&1
\end{pmatrix}
\begin{pmatrix}1&a_3&a_2\\&1&\\&&1
\end{pmatrix}\, ,\quad
k\in SO(3)\, .
$$ \label{foo}
Then instead of automorphizing $F(g;\lambda)$, one averages
over $\phi(\tau_1+i\tau_2)\, t_3^{\lambda_3-1}$ where $\phi$ is an 
$SL(2)$ cusp form.}.

Let us spell out this formula explicitly in a form
perhaps more familiar to physicists: Firstly observe that  the greatest common
divisor of any row (or column) of an element of $SL(3,\Zint)$ is unity. Let $m^\rt \equiv(m^1,m^2,m^3)$,
$n^\rt \equiv(n^1,n^2,n^3)$ where $\gcd(m^i)=\gcd(n^i)=1$. 
The summation matrix $\gamma$ is subject to the equivalence relation
\bea
&\gamma=
\begin{pmatrix}\, m\, &\, n\, &\, p\, \end{pmatrix}
\sim
\gamma n=
\begin{pmatrix}\, m\, &\, n\!+\!n_3m\, &\, p\!+\!n_1 n\! +\! n_2 m
\,\end{pmatrix}\, ,
&\nn\\
&n\equiv
\begin{pmatrix}1&n_3&n_2\\&1&n_1\\&&1\end{pmatrix}
\in N(\Zint)\, .&
\eea
Note that the column vector $p$ is completely determined modulo 
integer shifts by $m$ and $n$
once these two row vectors are specified. Therefore the summation
over $\gamma\in N(\Zint)
\backslash SL(3,\Zint)$ amounts to summing over $m^i$ and $n^i$
subject to $\gcd(m^i)=\gcd(n^i)=1$ and $n\sim n+n_3 m$, $n_3\in \Zint$.

To display the summand of~\eqn{Eisenstein} we compute the
Iwasawa decomposition of~$g\gamma$
\def\rt {{\rm t}}
\be
g\gamma=k'\ 
\begin{pmatrix}\sqrt{m^\rt  Gm}&*&*\\
&\sqrt{\frac{\ts m^\rt  Gm\  n^\rt  Gn- (m^\rt  G n)^2 }{\ts m^\rt  G m}} &*\\
&&
\frac{\ts 1}{\sqrt{\ts m^\rt  Gm\  n^\rt  Gn- (m^\rt  G n)^2 }}
\end{pmatrix}\, ,
\ee 
where $k'\in SO(3)$ and the matrix $G\equiv g^\rt g $.
Orchestrating, we find
\bea
\!\!\!E(g;\lambda)\!&=&\!\!\!
\sum_{
\begin{array}{c}
\ss\{ (m,n)\in \Zint^3\otimes \Zint^3 \ :\\
\ss \gcd(m)=\gcd(n)=1,\\
\ss m\neq 0,\, n\sim n+z m \neq 0\}
\end{array}}\!\!\!
{\Big(m^\rt  Gm\Big)^{-s}}\  
{\Big(m^\rt  Gm\  n^\rt  Gn- (m^\rt  G n)^2\Big)^{-t} }
\nn\\
&=&
\!\!\frac{1}{4\zeta(2s+2t)\zeta(2t)}\!\!\!\!\!
\sum_{
\begin{array}{c}
\ss\{ (m,n)\in \Zint^3\otimes \Zint^3 \ :\\
\ss m\neq 0,\, n\sim n+z m \neq 0\}
\end{array}}\!\!\!
{\Big(m^\rt  Gm\Big)^{-s}}\  
{\Big(m^\rt  Gm\  n^\rt  Gn- (m^\rt   G n)^2\Big)^{-t} }\, ,
\nn\\
\label{EisSl3}
\eea
where
\be
2s\equiv 1-\lambda_{32}\, ,\qquad
2t\equiv 1-\lambda_{21}\, .\qquad
\ee
In particular, observe that, in accordance with~\eqn{para_rep}, 
if we set $m^\rt =(1,x,v+xw)$ and $n^\rt =(0,1,w)$ (which makes sense
adelically)
the summand in~\eqn{EisSl3} matches the spherical vector~\eqn{spherical_A2}.

\subsection{Constant term computations}

\label{constant_term_app}

By construction, the Eisenstein series $E(g;\lambda)$ 
is invariant under right action of
$SL(3,\Zint)$ and therefore of its Borel
subgroup $N(\Zint)$, which acts on $(a_1,a_2,a_3)$ as
\be
E(a_1,a_2,a_3)=E(a_1+n_1,a_2+n_2+n_1a_3,a_3+n_3)\, ,\qquad
n_i\in \Zint\, .
\ee
(we drop the dependence on $t_i$ and $\lambda_i$).
This Heisenberg group includes two distinct $\Zint\times\Zint$
subgroups,
\be
N_1\equiv\left\{\begin{pmatrix}1&*&*\\&1&\\&&1\end{pmatrix}\right\}\, ,
\qquad
N_2\equiv\left\{\begin{pmatrix}1&&*\\&1&*\\&&1\end{pmatrix}\right\}\,
,
\ee
which can be viewed as the unipotent radicals of two 
associate maximal parabolic subgroups of $SL(3,\Zint)$,
\be
P_1\equiv
\left\{\begin{pmatrix}*&*&*\\{}&*&*\\&*&*\end{pmatrix}\right\}
\subset SL(3,\Real)\supset
\left\{\begin{pmatrix}*&*&*\\{}*&*&*\\&&*\end{pmatrix}\right\}
\equiv P_2\, .\label{parabolics}
\ee
One may therefore construct the two Fourier series expansions
\bea
E(a_1,a_2,a_3)&=&
\sum_{m_2,m_3}e^{2\pi i(m_2 a_2 + m_3 a_3)}E^{m_2,m_3}_{P_1}(a_1)
\label{F1}\\
&=&
\sum_{m_1,m_2}e^{2\pi i(m_1 a_1 + m_2 [a_2- a_1a_3])}E^{m_1,m_2}_{P_2}(a_3)\, .
\label{F2}
\eea
The Fourier zero-modes $E^{0,0}_{P_1}(a_1)\equiv E_{P_1}(g,\lambda)$
and  $E^{0,0}_{P_2}(a_3)\equiv E_{P_3}(g,\lambda)$, are called the
constant terms of the Eisenstein series $E(g;\lambda)$ with respect
to parabolics $P_1$, $P_2$, respectively. Evidently they may also
be obtained by performing compact integrations
\be
E_{P_{i}}(g,\lambda)
=\int_{N_{i} /[SL(3,\Zint)\cap N_{i}]}dn\  E(gn;\lambda)\, .
\ee
This is a general definition for constant terms. A third
constant term computation, relative to the minimal parabolic, is also 
possible
\be
E_P(g,\lambda)
=\int_{N/[SL(3,\Zint) \cap N]}dn\  E(gn;\lambda)\, .
\ee
The $N/[SL(3,\Zint) \cap N]$ is simply the set of upper
triangular matrices
\be
N/[SL(3,\Zint) \cap N]=
\left\{\begin{pmatrix}1&n_1&n_2\\&1&n_3\\&&1\end{pmatrix}:
0\leq n_1,n_2,n_3<1
\right\}\, .
\ee
Hence, from the Fourier expansions~\eqn{F1} and~\eqn{F2}
we find
\be
E_P(g;\lambda)=\int_0^1 dn_1 E^{0,0}_{P_1}(a_1+n_1)
=\int_0^1 dn_3 E^{0,0}_{P_2}(a_3+n_3)\, .
\label{check}
\ee
These relations provide a useful check on the constant term
computations.

Although constant term computations amount to the usual 
large volume expansion in String Theory--amenable to 
explicit computations via Schwinger's integral representation
and Poisson resummation, for general parabolics and higher
groups this method becomes unwieldy. 
Fortunately the task of computing $SL(3)$ constant terms has already
been tackled by mathematicians. Here we rely heavily on the analysis
of~\cite{Miller}.

First, recall the $SL(2)$ case
\bea
\sum_{\begin{array}{c}\ss (m,n)\neq 0,\\[-2mm]\ss \gcd(m,n)=1\end{array}}
\left(
\frac{\tau_2}{|m+n\tau|^2}
\right)^s
&=&\frac{1}{2\zeta(s)}\sum_{(m,n)\neq 0}
\left(
\frac{\tau_2}{|m+n\tau|^2}
\right)^s\nn\\
&=&\tau_2^s+\frac{\xi(2s-1)}{\xi(2s)}\ \tau_2^{1-s}\, 
+\sum_{n\neq 0} a_n(\tau_2)\  e^{2\pi i n\tau_1}.\nn\\
&\equiv& E_{1-2s}(\tau)\, .\label{sl2eisen}
\eea
Here 
\be
\xi(s)\equiv\pi^{-s/2}\Gamma(s/2)\zeta(s)\, ,
\ee is the completed
Riemann Zeta function and
\be
a_n(\tau_2)=\frac{\sqrt{\tau_2}}{\xi(2s)}\,
\mu_s(n)n^{s-1/2}K_{s-1/2}(2\pi\tau_2 |n|)\, ,
\qquad
\mu_s(n)\equiv \sum_{m|n}m^{-2s+1}\, .
\ee
Note that, in particular,
\be
\int_0^1 d\tau_1 E_s(\tau)=\tau_2^{\frac12-\frac1s}+
\frac{\xi(-s)}{\xi(s)}\, \tau_2^{\frac12+\frac1s}\, .
\label{Sl2}
\ee
Here $\xi(s)=\pi^{-s/2}\Gamma(s/2)\zeta(s)$ is the completed
Riemann Zeta function. Selberg's functional relation 
\be
\xi(s)E_{s}(\tau)=\xi(-s)E_{-s}(\tau)\, , \label{Selberg}
\ee
may be obtained
from the constant terms. Indeed, 
multiplying~\eqn{sl2eisen} by $\zeta(2s)$
and using $\xi(s)=\xi(1-s)$ we see that these are indeed invariant under
$s\rightarrow 1-s$. 
Note that the symmetric form of the $SL(2)$ Eisenstein series in~\eqn{Selberg}
is related to the usual one based on the fundamental representation
of $SL(2)$ by $E_{1-2s}(\tau)=E^{SL(2)}_{{\bf 2},s}/(2\zeta(2s))$.

For $SL(3)$ the following formulae hold:
\bea
E_{P_1}(g,\lambda)&=&t_3^{\lambda_{31}-\frac12\lambda_{21}-\frac32}\ 
E_{\lambda_{21}}(\tau)
+\frac{\xi(\lambda_{23})}{\xi(\lambda_{32})}\ 
t_3^{\lambda_{21}-\frac12\lambda_{31}-\frac32}
E_{\lambda_{31}}(\tau)\nn\\
&+&
\frac{\xi(\lambda_{12})\xi(\lambda_{13})}{\xi(\lambda_{21})\xi(\lambda_{31})}
\ t_3^{\lambda_{12}-\frac12\lambda_{32}-\frac32}
E_{\lambda_{32}}(\tau)\, ,
\label{p1}
\eea
where $\tau=a_1+i\frac{t_1}{t_2}$ in this formula;
\bea
E_{P_2}(g,\lambda)&=&t_1^{\lambda_{12}-\frac12\lambda_{32}+\frac32}\ 
E_{\lambda_{32}}(\tau)
+\frac{\xi(\lambda_{12})}{\xi(\lambda_{21})}
t_1^{\lambda_{21}-\frac12\lambda_{31}+\frac32}
E_{\lambda_{31}}(\tau)\nn\\
&+&
\frac{\xi(\lambda_{23})\xi(\lambda_{13})}{\xi(\lambda_{32})\xi(\lambda_{31})}
t_1^{\lambda_{31}-\frac12\lambda_{21}+\frac32}
E_{\lambda_{21}}(\tau)\, ,
\label{p2}
\eea
here $\tau=a_3+i\frac{t_2}{t_3}$.

By virtue of~\eqn{check}, inserting the constant term result for 
the $SL(2)$ Eisenstein series~\eqn{Sl2} in either~\eqn{F1}
or~\eqn{F2}, yields
\be
\begin{array}{ccccc}
E_{P}(g,\lambda)&=&
 t_1^{\lambda_1+1}t_2^{\lambda_2}t_3^{\lambda_3-1}
&+&
 t_1^{\lambda_2+1}t_2^{\lambda_1}t_3^{\lambda_3-1}\
 \frac{\xi(\lambda_{12})}{\xi(\lambda_{21})}
\\[3mm]&&
 t_1^{\lambda_1+1}t_2^{\lambda_3}t_3^{\lambda_2-1}\
 \frac{\xi(\lambda_{23})}{\xi(\lambda_{32})}
&+&
 t_1^{\lambda_3+1}t_2^{\lambda_1}t_3^{\lambda_2-1}\
 \frac{\xi(\lambda_{13})}{\xi(\lambda_{31})}
 \frac{\xi(\lambda_{23})}{\xi(\lambda_{32})}
\\[3mm]&&
 t_1^{\lambda_2+1}t_2^{\lambda_3}t_3^{\lambda_1-1}\
 \frac{\xi(\lambda_{12})}{\xi(\lambda_{21})}
 \frac{\xi(\lambda_{13})}{\xi(\lambda_{31})}
&+&\, 
 t_1^{\lambda_3+1}t_2^{\lambda_2}t_3^{\lambda_1-1}\
 \frac{\xi(\lambda_{12})}{\xi(\lambda_{21})}
 \frac{\xi(\lambda_{13})}{\xi(\lambda_{31})}
 \frac{\xi(\lambda_{23})}{\xi(\lambda_{32})}
\\[5mm]&=&
{\ts \sum_{s\in S(3)}}t_1^{\lambda_{s(1)}+1}t_2^{\lambda_{s(2)}}
t_3^{\lambda_{s(3)}-1}&&\hspace{-1.9cm}
\prod_{\!\!\!\!\!\begin{array}{c}\ss 1\leq i<j\leq 3\\[-1mm]\ss
 s^{-1}(i)>s^{-1}(j)\end{array}}
\frac{\xi(\lambda_{ij})}{\xi(\lambda_{ji})}\, .\hspace{1cm}
\end{array}
\ee
In particular observe that
\bea
\label{selb3}
\xi(\lambda_{21})\xi(\lambda_{31})\xi(\lambda_{32})\
E_P(g,\lambda)=\hspace{7cm}\nn\\[2mm]
\qquad\qquad
\sum_{s\in S(3)}\ t_1^{\lambda_{s(1)}+1}t_2^{\lambda_{s(2)}}
t_3^{\lambda_{s(3)}-1}\ 
\xi(\lambda_{s(2)s(1)})\xi(\lambda_{s(3)s(1)})\xi(\lambda_{s(3)s(2)})\, ,
\eea
is manifestly symmetric under the action of the Weyl group $S(3)$ on
$\lambda$. This relation in fact extends to the
Eisenstein series $E(g,\lambda)$ and 
is the generalization of the Selberg functional
relation~\eqn{Selberg} to $SL(3)$. 

Another interesting calculation is to compute the constant terms in
the case $2t=1-\lambda_{21}=0$, {\it i.e.} when only the quadratic
summand is present in~\eqn{EisSl3} which ought correspond
to the fundamental representation Eisenstein series
$E^{SL(3)}_{{\bf 3},s}(g)$. Setting $t=0$ in~\eqn{EisSl3}
leads to a divergent sum over $n$ but the analytic continuation
to $t=0$ is regular and can be obtained by studying the constant
term formulae:
\bea
E_{P_1}(g;1-2s,1)&=&
t_3^{-2s}+t_3^{s-3/2}\ \frac{\xi(2s-1)}{\xi(2s)}\ 
E_{2-2s}(\tau)\nn\\
E_{P_2}(g;1-2s,1)&=&
t_1^{-2s+3}\ \frac{\xi(2s-2)}{\xi(2s)}+t_1^s\ E_{1-2s}(\tau)\, .
\label{constantterm}
\eea
Here we used $E_1(\tau)=1$ and the fact that $\xi(s)$ has a
pole at $s=1$.
Indeed these results agree with those for the $SL(d+1)$ Eisenstein series in
the fundamental representation defined by
\be
\overline E^{SL(d+1)}_{{\bf d+1},s}(g)\equiv
\sum_{\begin{array}{c}\ss m\in {\mathbb Z}^{d+1}\setminus \{0\}
\\[-2mm] \ss \gcd(m)=1\end{array}}
\left(\frac{1}{mG m^\rt }\right)^s\equiv 
\frac{1}{2\zeta(2s)}E^{SL(d+1)}_{{\bf d+1},s}(g)\, ,
\ee
for which a simple large volume computation yields
\bea
\int_0^1 d^dn\  \overline E^{SL(d+1)}_{{\bf d+1},s}(gn)=
\left\{
\begin{array}{ll}
t_{d+1}^{-2s}+\frac{\xi(2s-1)}{\xi(2s)}\ 
t_{d+1}^{\frac{2s-1}{d}-1}\overline E^{SL(d)}_{{\bf d},s-1/2}(\gamma)\, ,&\;\; 
n=
\scalebox{.6}{$
\begin{pmatrix}1&n_1&\cdots&n_d\\&1&&\\&&\ddots&\\&&&1\end{pmatrix}
$}\, ,
\\ \\
t_{1}^{1+d-2s}\ \frac{ \xi(2s-d)}{\xi(2s)}+
t_{1}^{2s/d}\ \overline E^{SL(d)}_{{\bf d},s}(\gamma)\, , &\;\;
n=
\scalebox{.6}{$
\begin{pmatrix}1&&&n_1\\&\, 1\,&&n_2\\&&\; \ddots\; &\vdots\\&&&1\end{pmatrix}
$}
\, .
\end{array}
\right.\nn
\eea
\be\ee
Note that $\gamma\in SL(d)$ is obtained by respectively 
decomposing $g=a\gamma n$ with
$n\in N_{1,d}$ and $a\in M_{1,d}$. In the case $d=2$,  
$\overline E^{SL(2)}_{{\bf 2},s}(\gamma)=E_{1-2s}(\tau)$ with $\tau$ as
in footnote \ref{foo}.

\section{Membrane world-volume integrals}
\label{wv_app}

An alternative proposal to obtain an $SL(2)$ automorphic
form from an $SL(3)$ one, is to integrate
over a volume modulus by decomposing $SL(3)_{NL}
\supset SL(2)_{U}\times {\mathbb R}^+$. Here we perform this computation
for a simple toy model--an $SL(3)$ Eisenstein series in the fundamental 
representation $E_{{\bf 3},s}^{SL(3)}(g_{NL})$.
Firstly, we decompose the $SL(3)$ moduli as
\be
g_{NL}=
\left(\begin{array}{c|c}\nu g_U \ &\ 0
\\\hline 0 \ & \ \frac{1}{\nu^2}\end{array}\right)\, .
\ee
Holding the $2\times2$ metric $g_U=g_U(\tau)$ constant, we can compute
the metric on the membrane world-volume modulus $d\sigma^2=-\frac16\tr\, dg
dg^{-1}=\frac{d\nu^2}{\nu^2}$ to determine the measure of
integration. Therefore we study 
\bea
\int_0^\infty \frac{d\nu}{\nu}\ 
E^{SL(3)}_{{\bf 3},s}(g)&=&\int_0^\infty \frac{d\nu}{\nu}
\sum_{
\stackrel{\{(\mu,m):\mu\in{\mathbb Z}^2\, ,\, m\in{\mathbb Z}}
{\sss (\mu,m)\neq (0,0,0)\}} }
\left(\frac{1}{\nu \mu^\rt  \gamma \mu + \frac{m^2}{\nu^2}}\right)^s\\
&=&
2\ G(s)\ \zeta\Big(\frac{2s}{3}\Big)\ 
E^{SL(2)}_{{\bf 2},\frac{2s}{3}}(g_U)+{\cal A}\, ,
\eea
where 
\be
G(s)\equiv\int_0^\infty
\frac{d\nu}{\nu^{1-2s}(1+\nu^3)^s}=
\frac{\Gamma(\frac{s}{3})\Gamma(\frac{2s}{3})}{3\Gamma(s)}
\, .
\ee
The complete integral is however divergent, because of the degenerate
contributions at $m=0$ or $\mu=(0,0)$, which are given by the formal 
expression
\be
{\cal A} = \int_0^\infty \frac{d\nu}{\nu^{1+s}}\, \Big(\frac12+
E_{{\bf2},s}^{SL(2)}(g_U)\Big)\, .
\ee
This supports the conclusion in the Text that the membrane volume
integration is divergent. We have not been able to find a consistent
regularization scheme.

\end{appendix}

\providecommand{\href}[2]{#2}
\begingroup
\raggedright
\endgroup
\end{document}